**Title Page**

Article Title: Human movement reconstruction through latent structural representations during assisted transitions and multidirectional hopping

Authors and affiliations:
Guilherme Manna Cesar[1], Ligia Yumi Mochida[1], Chady Ghnatios[2]
[1] Department of Physical Therapy, Brooks College of Health, University of North Florida.
[2] Department of Mechanical Engineering, Advanced Manufacturing & Materials Innovation program, College of Computing, Engineering & Construction, University of North Florida.

Corresponding author information: Guilherme Manna Cesar, PhD, PT.
Department of Physical Therapy, Brooks College of Health, University of North Florida.
1 UNF Drive, Jacksonville, Florida 32224-7699.
Email: g.cesar@unf.edu
Phone / Fax: 904.620-2664 / 2251

## Abstract

The quantification of three-dimensional human kinematics is fundamental to neurorehabilitation and musculoskeletal research, yet atypical posture, physical assistance, and rapid movement create conditions of partial observability. Here, we evaluate a latent-structure-informed framework combining convolutional rank-reduction autoencoding with feedforward regression to reconstruct movement from synchronized video. Three case-based demonstrations span therapist-assisted sit-to-stand in a child with cerebral palsy, unassisted sit-to-stand in a typically developing child, and multidirectional single-leg hopping in a young adult. Reconstruction was examined through positional trajectories, projected angular descriptors, and hip-knee coupling. Across test repetitions excluded from model training, mean absolute errors in a predicted marker position during assisted sit-to-stand ranged from 12.02 to 57.11 mm. Sagittal knee and trunk descriptor errors were 5.32° and 5.38° in the typically developing case. Hip-knee cyclograms captured aspects of coupled-motion patterns in both cases, with closer trajectory correspondence in the typically developing case. Sixteen test hops yielded a vertical knee-marker mean absolute error of 15.58 mm. These findings connect latent-structure-informed reconstruction with biomechanically interpretable movement features, providing a foundation for video-derived assessment spanning clinically atypical movement and high-dynamic athletic movement.

## Introduction

The quantification of high-dimensional biological systems is frequently impeded by partial observability, where the underlying state of a process must be inferred from incomplete or corrupted data streams. In human movement science, three-dimensional (3D) kinematics are fundamental to musculoskeletal research, neurorehabilitation, and clinical decision-making, and are commonly quantified through external markers placed on the skin and tracked by specialized laboratory systems.[1–3] These protocols provide detailed information on marker positions and model-derived joint and segment kinematics (Fig. 1), but require precise marker placement, specialized equipment, and sufficient marker visibility.[1,4] The protocols' sensory and procedural demands may complicate participation for children with neurological disabilities and sensory-processing abnormalities.[5,6] Beyond clinical contexts, rapid segment reorientation and self-occlusion can interrupt motion-capture recordings, and the characteristics of missing data can affect subsequent biomechanical calculations.[7,8] Such constraints motivate reconstruction approaches that relate the available visual information to the configuration of an articulated body, rather than evaluating landmark visibility in isolation.

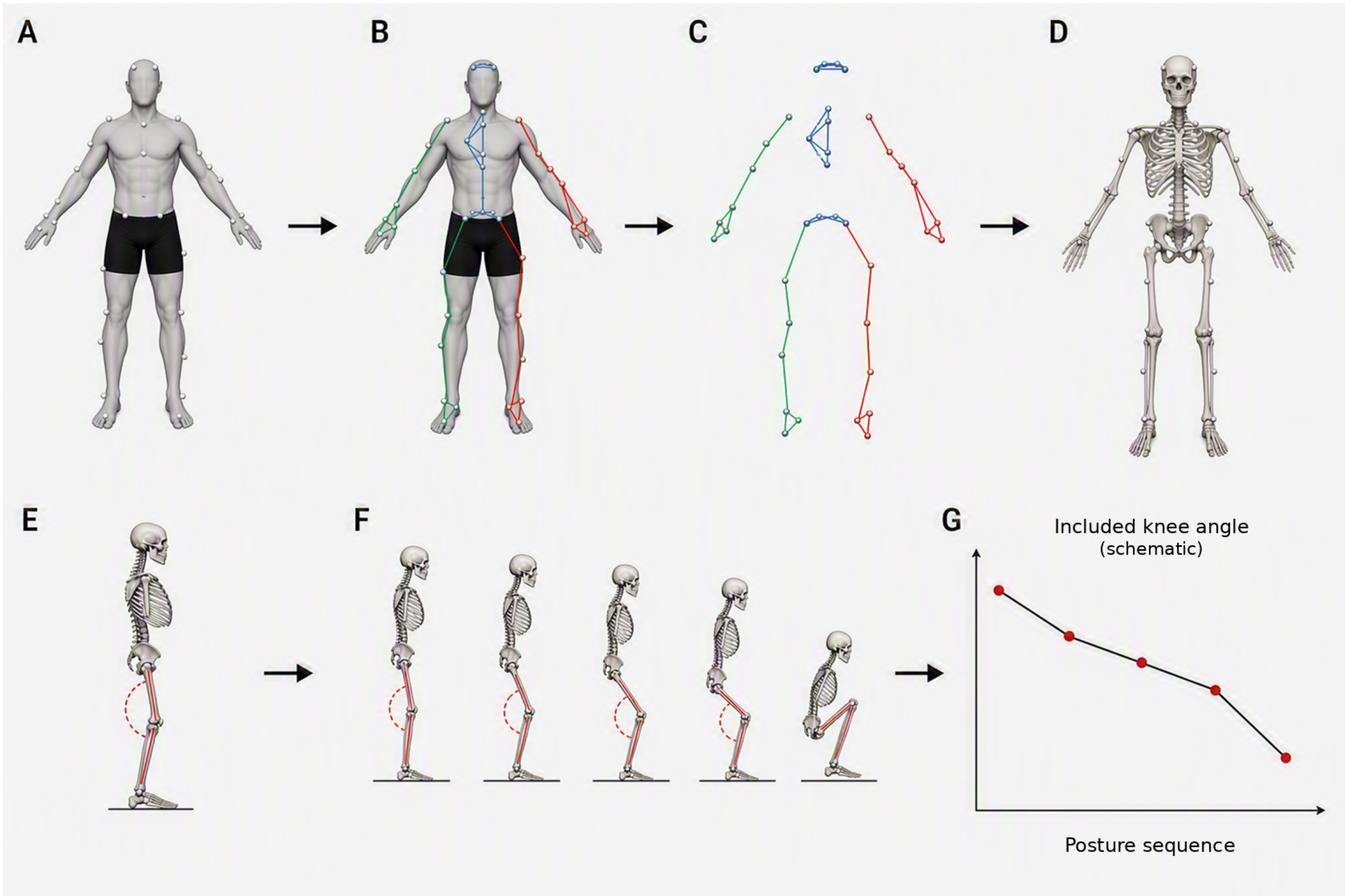

**Fig. 1 | From surface-marker observations to model-derived human kinematics.** a, Reflective markers positioned on an illustrative human model. b, Marker positions associated with body regions. c, The corresponding marker configuration. d, A skeletal representation used to illustrate the relationship between surface observations and biomechanical modeling. e, An included angle between thigh and shank. f, A sequence of increasingly flexed postures. g, The corresponding schematic reduction in the included angle. Panels e–g are schematic illustrations of the measurement principle rather than recordings from the study. The included angle decreases with increasing knee flexion.

Markerless motion capture has emerged as a scalable alternative to these protocol-driven constraints.[4] Deep neural networks can estimate clinically relevant gait metrics from single-camera videos, and pixel-based machine learning can recover selected spatiotemporal gait parameters against instrumented-walkway measurements.[9,10] Open-source methods such as DeepLabCut and OpenPose have further expanded access to video-based landmark estimation.[11,12] However, accurate recovery of individual movement deviations remains a challenge. In children with cerebral palsy, a multiview markerless system tracked several gait features but did not consistently capture deviations in pelvic tilt and transverse hip rotation.[13] Our prior multimodal evaluation similarly showed that two-dimensional markerless hip-knee cyclograms could approximate expected movement patterns while diverging when pediatric sit-to-stand involved more pronounced out-of-plane motion.[6] These findings identify a biomechanical problem beyond landmark detection: Whether video-derived trajectories retain the relationships among body segments that distinguish individual movement strategies.

The recovery of these relationships is particularly relevant during functional transitions that combine lower-extremity motion with control of the trunk. Among these transitions, sit-to-stand supports transfers, walking initiation, and self-care, and its performance in children with cerebral palsy is associated with age and gross motor function.[14,15] In this population, impaired selective motor control further motivates assessment of how joints move together rather than describing each trajectory in isolation.[16] Hip-knee cyclograms offer a direct representation of this coupling and have been used to examine pediatric sit-to-stand and to characterize relationships with gross motor and gait function during walking.[6,17] These representations provide complementary information: positional trajectories describe the spatial progression of movement, angular descriptors express relationships among body segments, and cyclograms describe how paired angular quantities evolve together. Examining these levels jointly therefore allows reconstruction to be assessed in terms of the movement organization retained, rather than positional error alone. To extend this question beyond pediatric transitions, multidirectional single-leg hopping provides a complementary movement regime involving repeated propulsion, flight, landing, and redirection.

We here evaluate a framework that combines nonlinear model order reduction with feedforward regression to reconstruct movement from synchronized video (Fig. 2). Rank-Reduction autoencoders constrain the effective dimension of the latent representation through truncated singular value decomposition, defining effective latent dimensionality through retained rank rather than the dimensionality of the network architecture.[18–20] View-specific latent representations are combined to predict marker coordinates, from which relationships among body segments are calculated. Here, structure refers to the low-rank organization of the learned representation rather than an explicitly encoded anatomical graph. This rationale is consistent with evidence that behavioral constraints can improve learning-based kinematic estimation and that coupled visual and kinematic manifolds can support movement tracking.[21,22] The computational strategy examined here couples rank-constrained representation learning with marker-coordinate regression. Its biomechanical consequences are examined through positional, projected angular, and coupled-motion readouts across contrasting demands.

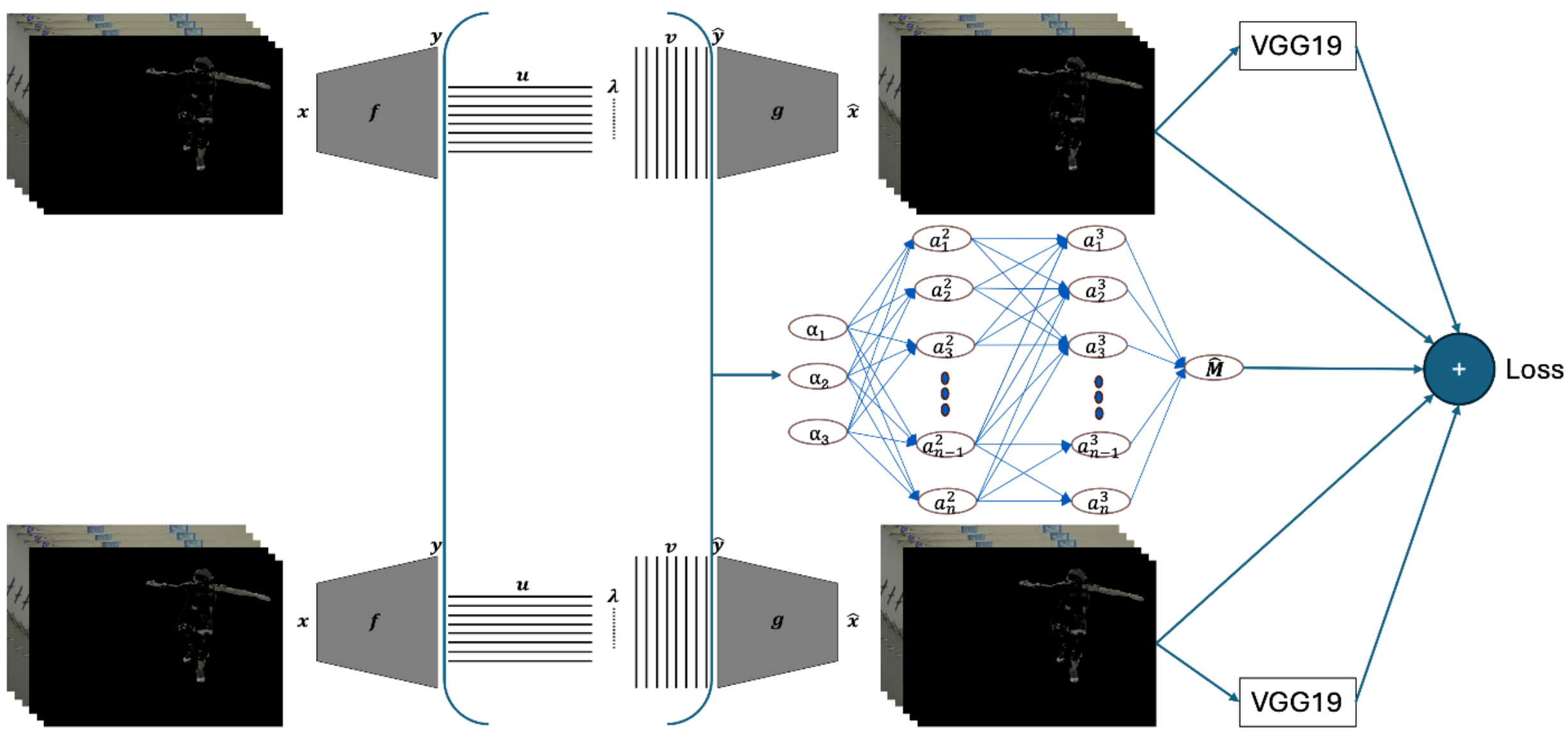

**Fig. 2 | Two-view rank-reduction autoencoder framework for video-derived marker reconstruction.** Synchronized image representations from two camera views are processed by separate convolutional encoders (*f*). Singular value decomposition constrains the rank of the latent representations, which support image reconstruction through view-specific decoders (*g*) and are combined as inputs to a feedforward multilayer perceptron for marker-coordinate prediction. The image inputs and reconstructions are denoted by *$\mathbf{x}$* and *$\hat{\mathbf{x}}$*, respectively, and the encoded and rank-constrained representations by *$\mathbf{y}$* and *$\hat{\mathbf{y}}$*. The singular value decomposition (SVD) in the latent space reproduces the left singular vectors $\boldsymbol{u}$, the right singular vectors $\boldsymbol{v}$ and the singular values $\lambda$. The truncation is performed based on the reduction of the $\lambda$ values, and the reconstruction of the quantity $\boldsymbol{\alpha} = \boldsymbol{u}^T \lambda\, \boldsymbol{v}$ is performed using the truncated SVD. The concatenation of $\boldsymbol{\alpha}$ from different views is passed into an MLP to predict the measures

outputs $\boldsymbol{M}$. The image-reconstruction branches contribute pixel-reconstruction and VGG19-feature losses, while the regression branch contributes the marker-coordinate loss. These losses are evaluated against their corresponding input-image or measured-marker targets during joint training; the reference targets are omitted from the schematic. Network nodes and connections are illustrative rather than an exact depiction of layer dimensions. Architecture details are provided in Supplementary Fig. 8.

The framework was evaluated across three case-based demonstrations selected to represent contrasting observational and movement demands: therapist-assisted sit-to-stand in a child with cerebral palsy, unassisted sit-to-stand in a typically developing child, and multidirectional single-leg hopping in a young adult. Evaluation progressed from positional reconstruction during the assisted transition to angular descriptors in the unassisted case and hip-knee coupling in both children. Hopping provided a complementary test of positional reconstruction in a distinct movement regime. Synchronized marker-based measurements provided the reference for assessing which movement features were recovered in held-out repetitions.

**Results**

Within the rank-reduction component, effective latent dimensionality is defined by the retained rank of the latent matrix through truncated singular value decomposition rather than latent-layer dimensionality alone.[18–20] During inference, the RRAE formulation projects newly encoded observations onto a basis identified from the training data.[18,19] In the present study, view-specific latent representations support image reconstruction and are concatenated as inputs to a feedforward multilayer perceptron for marker-coordinate prediction; the networks are jointly trained using image-reconstruction, perceptual, and marker-coordinate losses (Fig. 2). The resulting reconstructions are examined through positional trajectories, projected angular descriptors, hip-knee coupling, and multidirectional hopping.

While training was performed on 5 participants, the test outcomes presented here were obtained from recordings of three participants, with repetitions analyzed within each case rather than as independent participants. The test analyses comprised four assisted sit-to-stand repetitions, four unassisted sit-to-stand repetitions for the angular descriptors, three unassisted repetitions for the hip-knee cyclogram, and 16 individual hops. The analysis-specific counts and training summaries are provided in Supplementary Tables 1–3.

**Assisted sit-to-stand reconstruction differs across spatial components**

The framework was first evaluated during therapist-assisted sit-to-stand performed by the child with cerebral palsy, Gross Motor Function Classification System level III.[23] Across four test repetitions, mean absolute errors for the right-back marker trajectory were 27.00, 59.47, and 17.76 mm along the recorded X, Y, and Z axes, respectively (Fig. 3). Corresponding $R^2$ values were 0.564, -0.164, and 0.835. Thus, the Z-coordinate trajectory showed the strongest agreement among the three components, whereas the Y-coordinate reconstruction preserved the overall temporal progression and trajectory pattern despite lower positional agreement with the reference. The reconstruction therefore captured key features of the assisted transition, with fidelity varying across spatial components.

**Figure 3a | RBAK X**

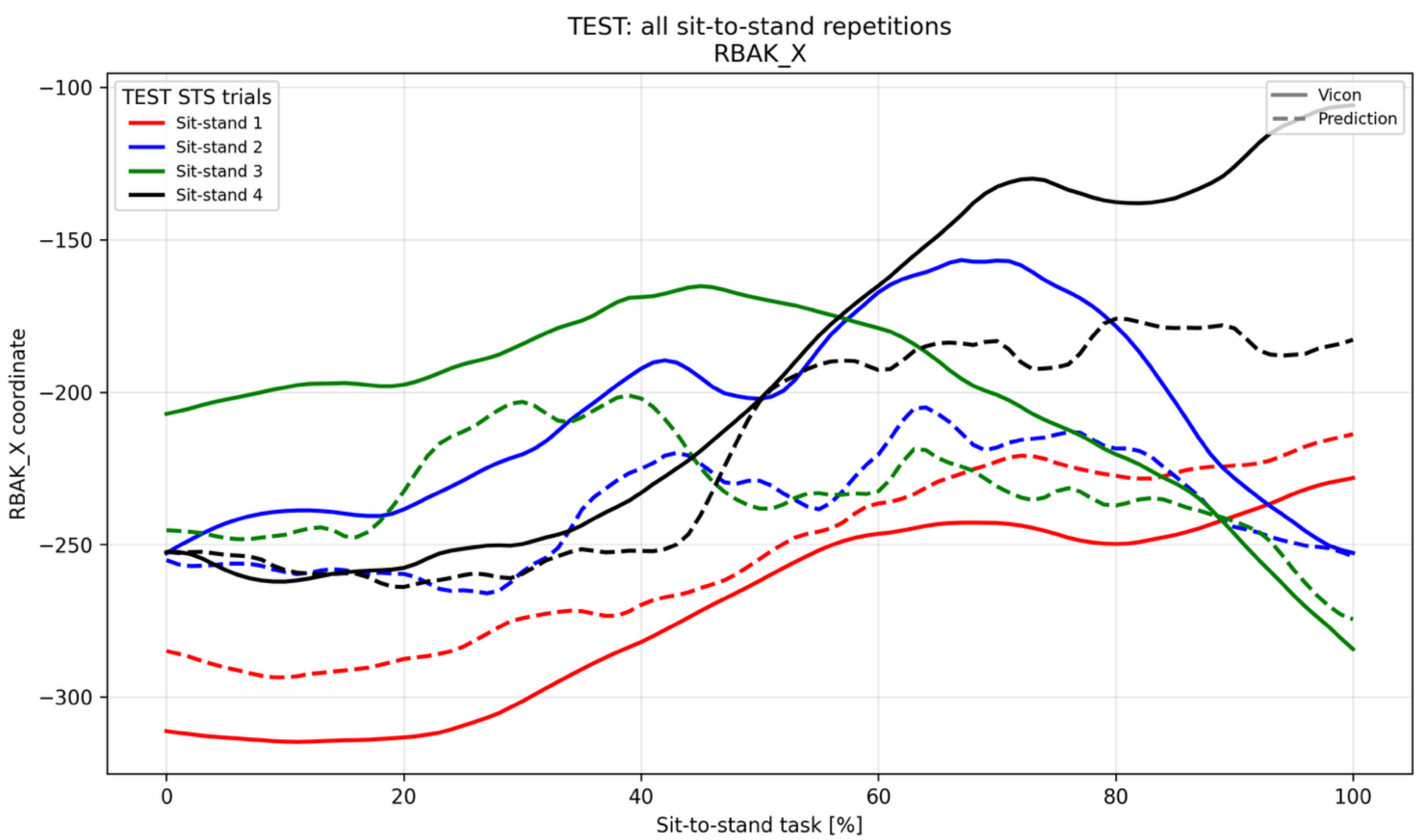

**Figure 3b | RBAK Y**

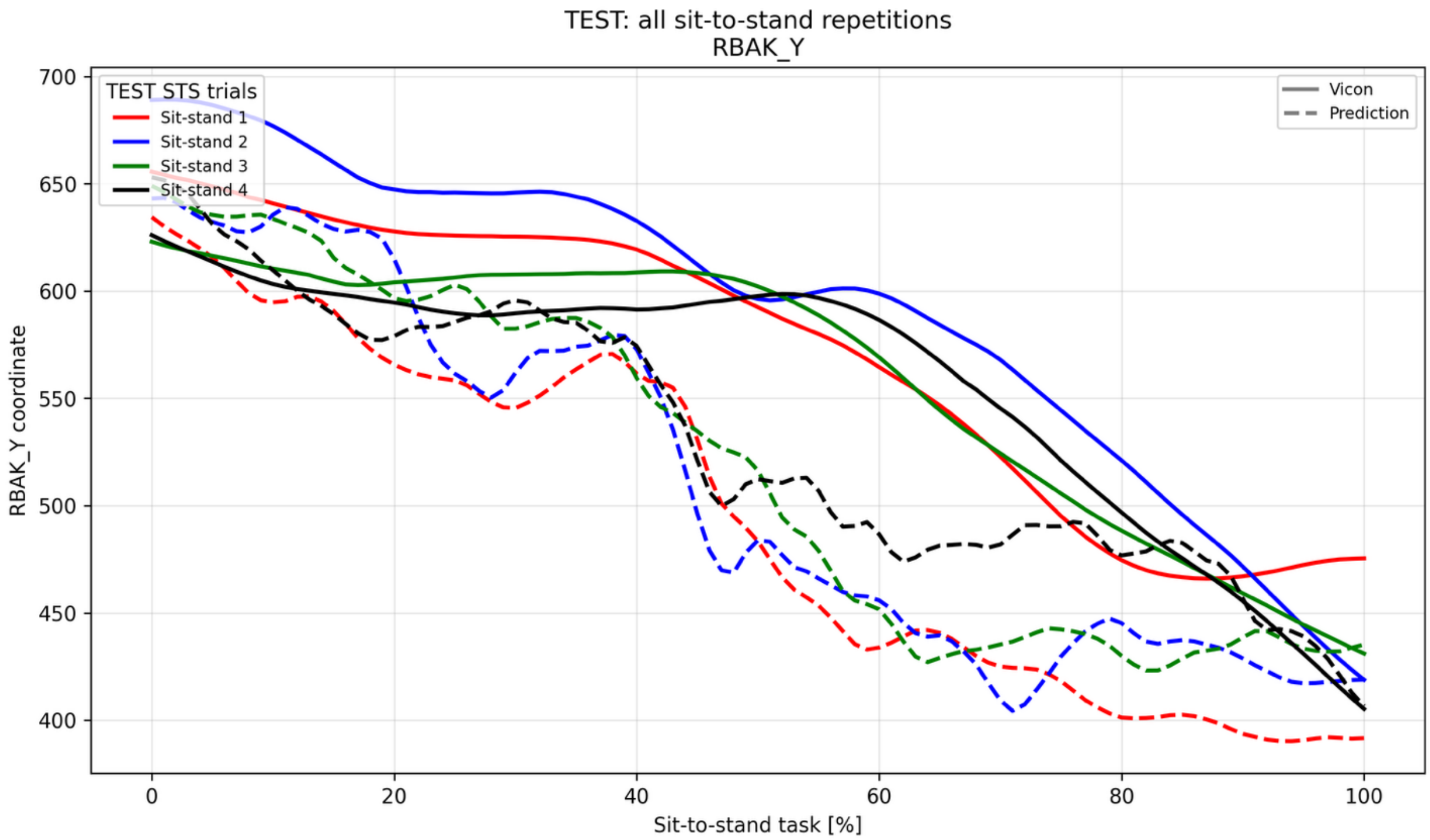

**Figure 3c | RBAK Z**

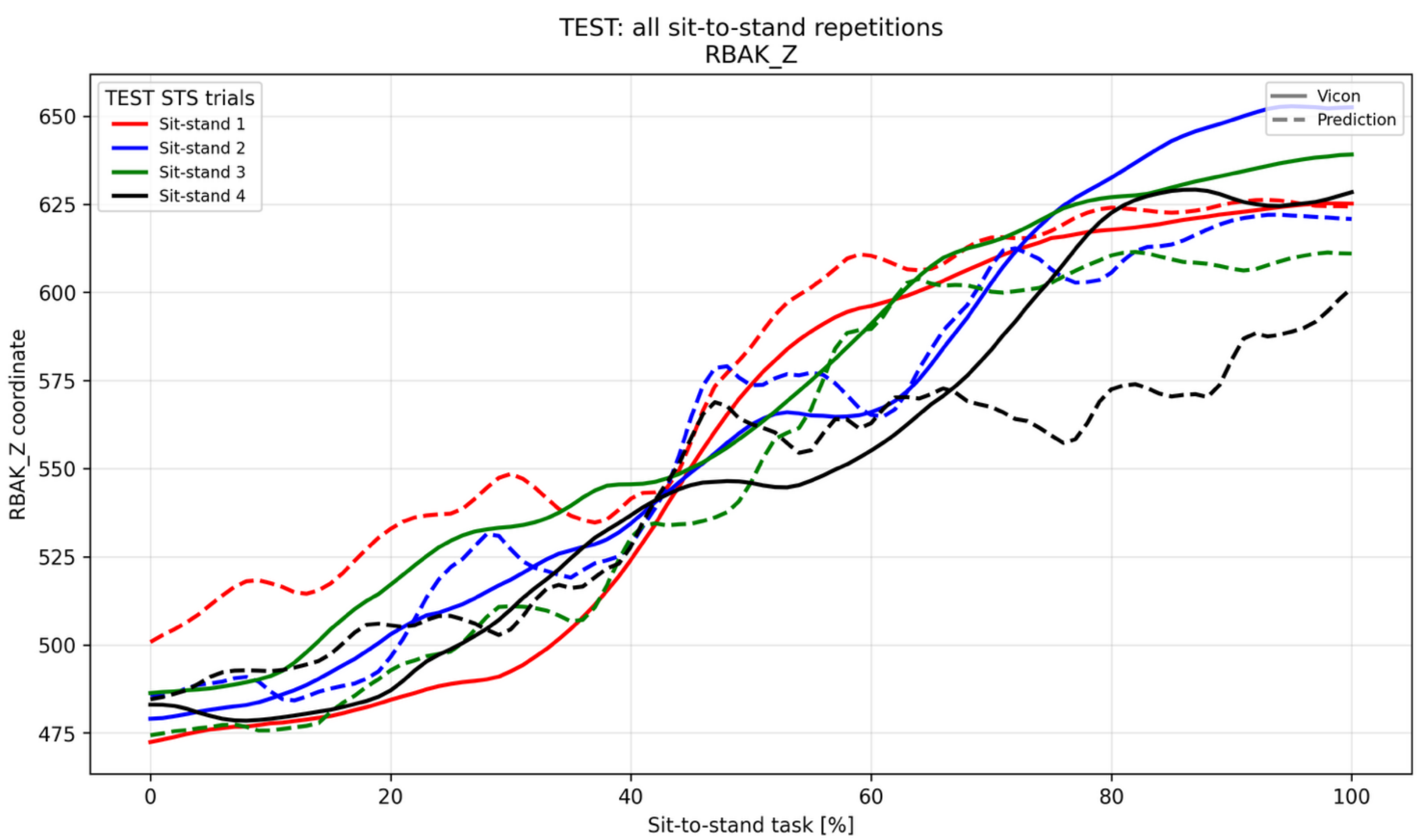

**Fig. 3 | Test reconstruction during therapist-assisted sit-to-stand.** a–c, Right-back marker (RBAK) coordinates along the recorded X, Y, and Z axes for four repetitions from the child with cerebral palsy. Solid curves indicate the synchronized marker-based reference, and dashed curves

indicate the corresponding predictions. Matching colors identify the four repetitions. The X, Y, and Z labels refer to the acquisition coordinate system; they are not anatomical planes. Coordinate-error summaries are reported in the Results and Supplementary Table 3. n = 4 repetitions from one participant.

**Unassisted sit-to-stand reconstruction is more consistent for sagittal descriptors**

The typically developing comparator was used to examine angular reconstruction without examiner assistance under the same acquisition configuration. Across four test repetitions, mean absolute errors for the sagittal-, frontal-, and transverse-plane projected thigh-shank descriptors were 5.65°, 4.46°, and 14.55°, respectively, with $R^2$ values of 0.944, 0.425, and 0.872 (Fig. 4a, c, e). The sagittal reconstruction reproduced the dominant progression toward knee extension, while the frontal-plane descriptor had the smallest absolute error among the three knee descriptors. The transverse-plane reconstruction retained the overall movement pattern with greater absolute angular error. In the frontal-plane trajectory (Fig. 4c), one repetition (Exp. 3–STS 3) contained a transient excursion in the marker-based reference that was not reproduced to the same extent by the processed prediction. The reported MAE and $R^2$ include this interval and therefore quantify correspondence with the recorded reference trajectory.

Pelvis-relative trunk reconstruction also varied across planes (Fig. 4b, d, f). The sagittal descriptor showed a mean absolute error of 5.38° and an $R^2$ of 0.818. The frontal-plane descriptor had a mean absolute error of 4.10° and an $R^2$ of −0.659; one repetition (Exp. 3–STS 1) contained a transient excursion in the marker-based reference that was not reproduced to the same extent by the processed prediction. The reported frontal-plane agreement statistics include this portion of the reference trajectory. The transverse-plane descriptor showed a mean absolute error of 3.45° and an $R^2$ of −1.280. Across these descriptors, mean absolute error ranged from 3.45° to 5.38°, with the strongest trajectory-level agreement observed in the sagittal plane.

## Figure 4a | Projected knee descriptor, sagittal

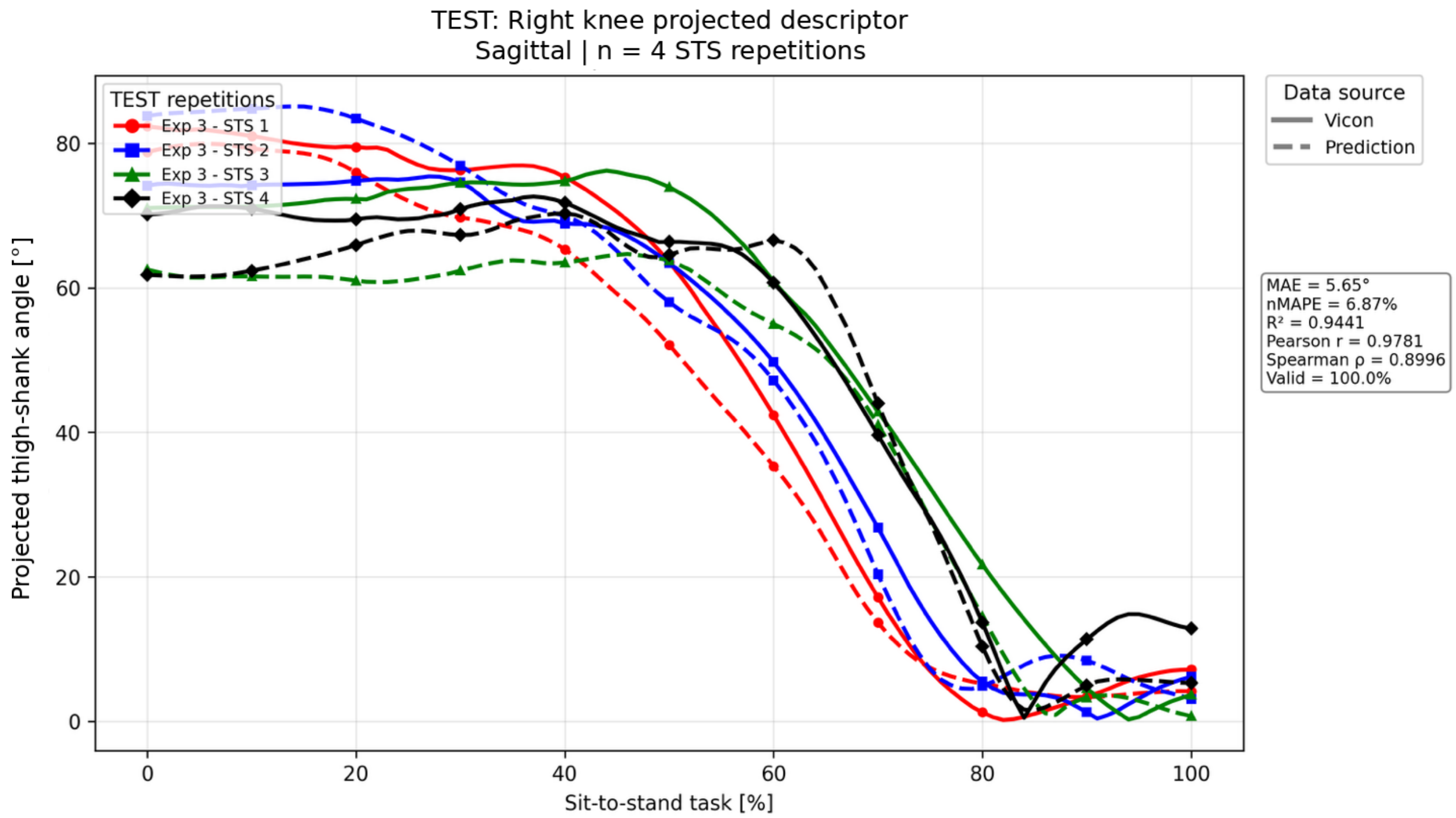

## Figure 4b | Trunk, sagittal

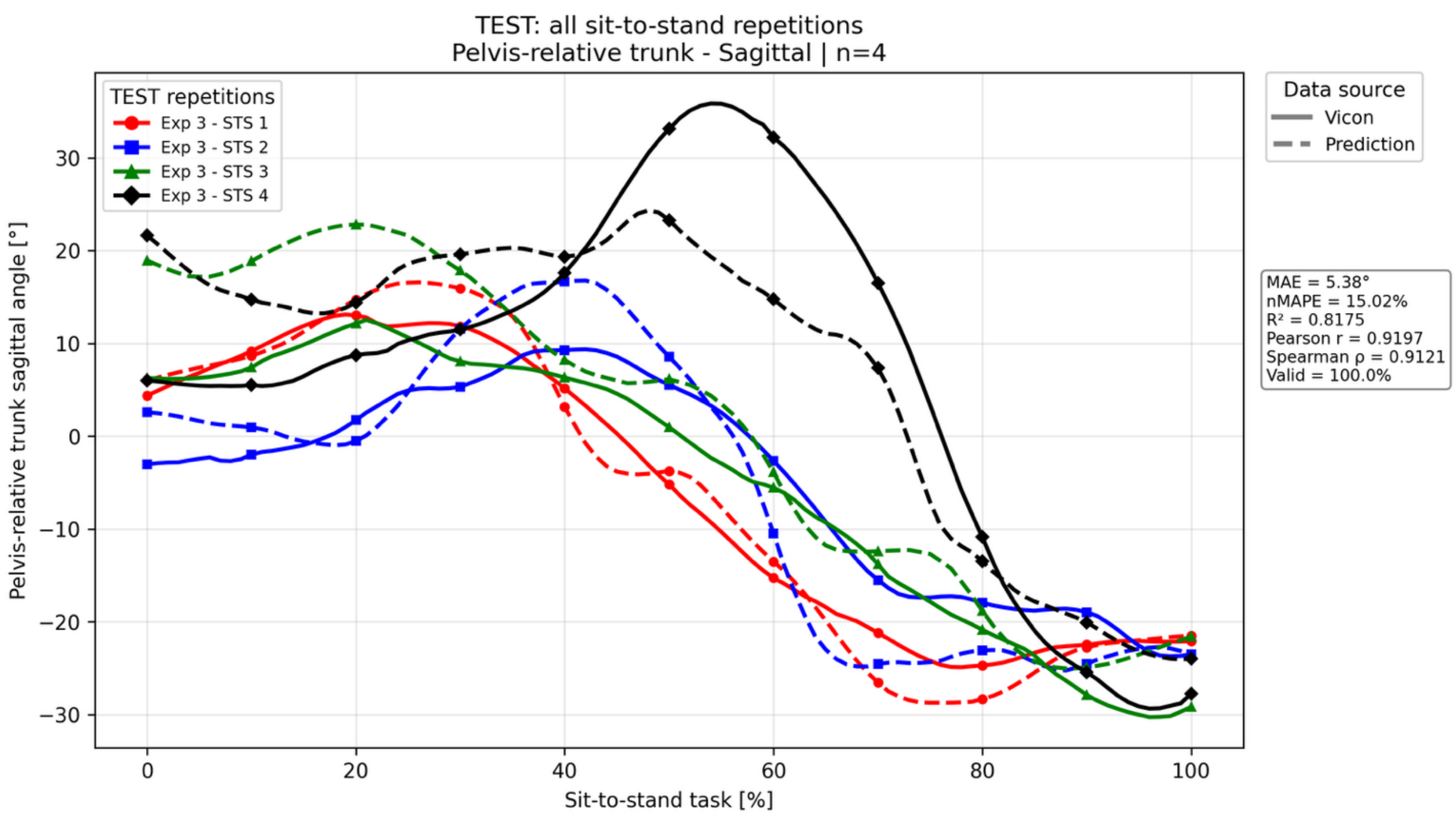

**Figure 4c | Projected knee descriptor, frontal**

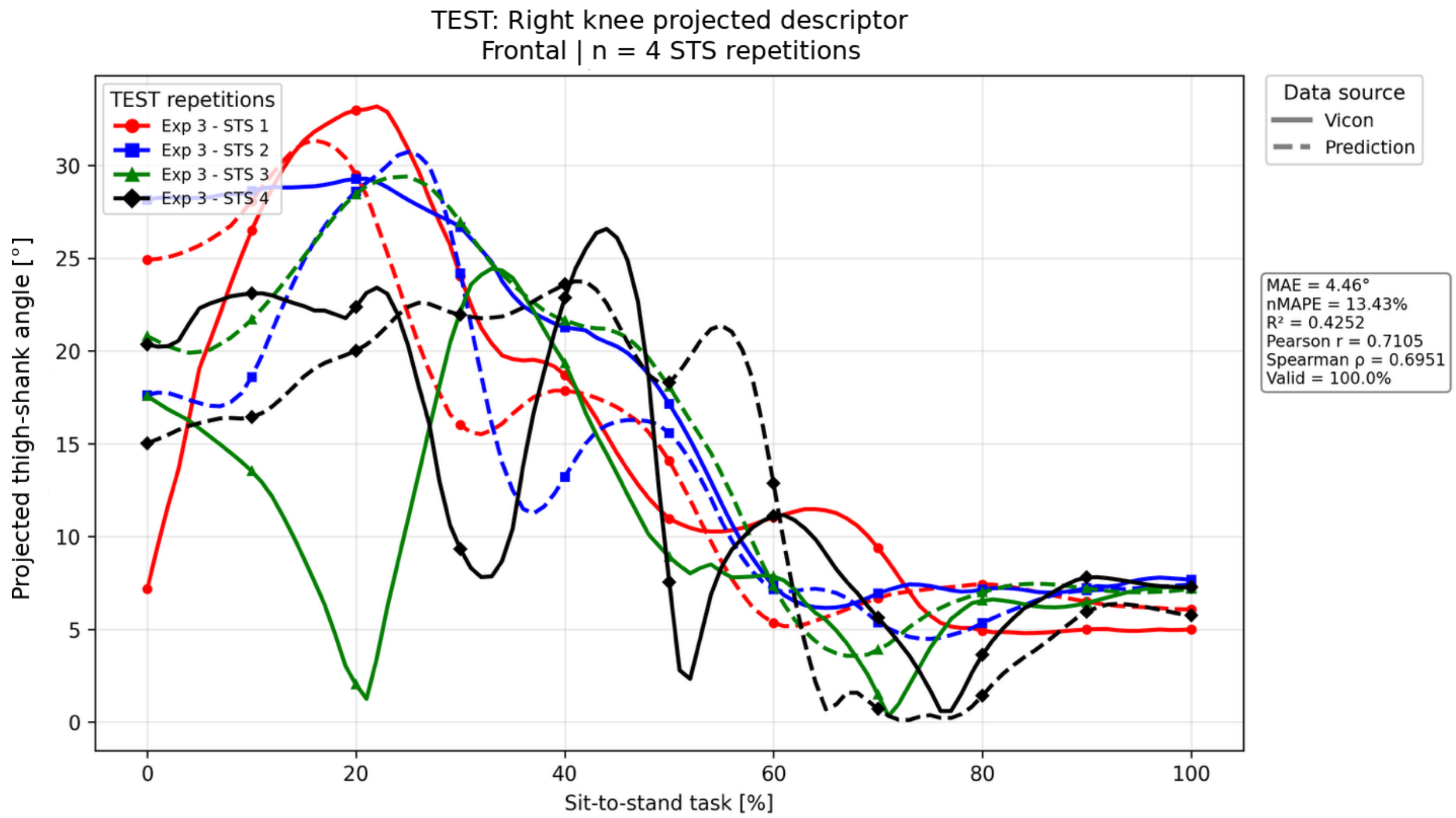

**Figure 4d | Trunk, frontal**

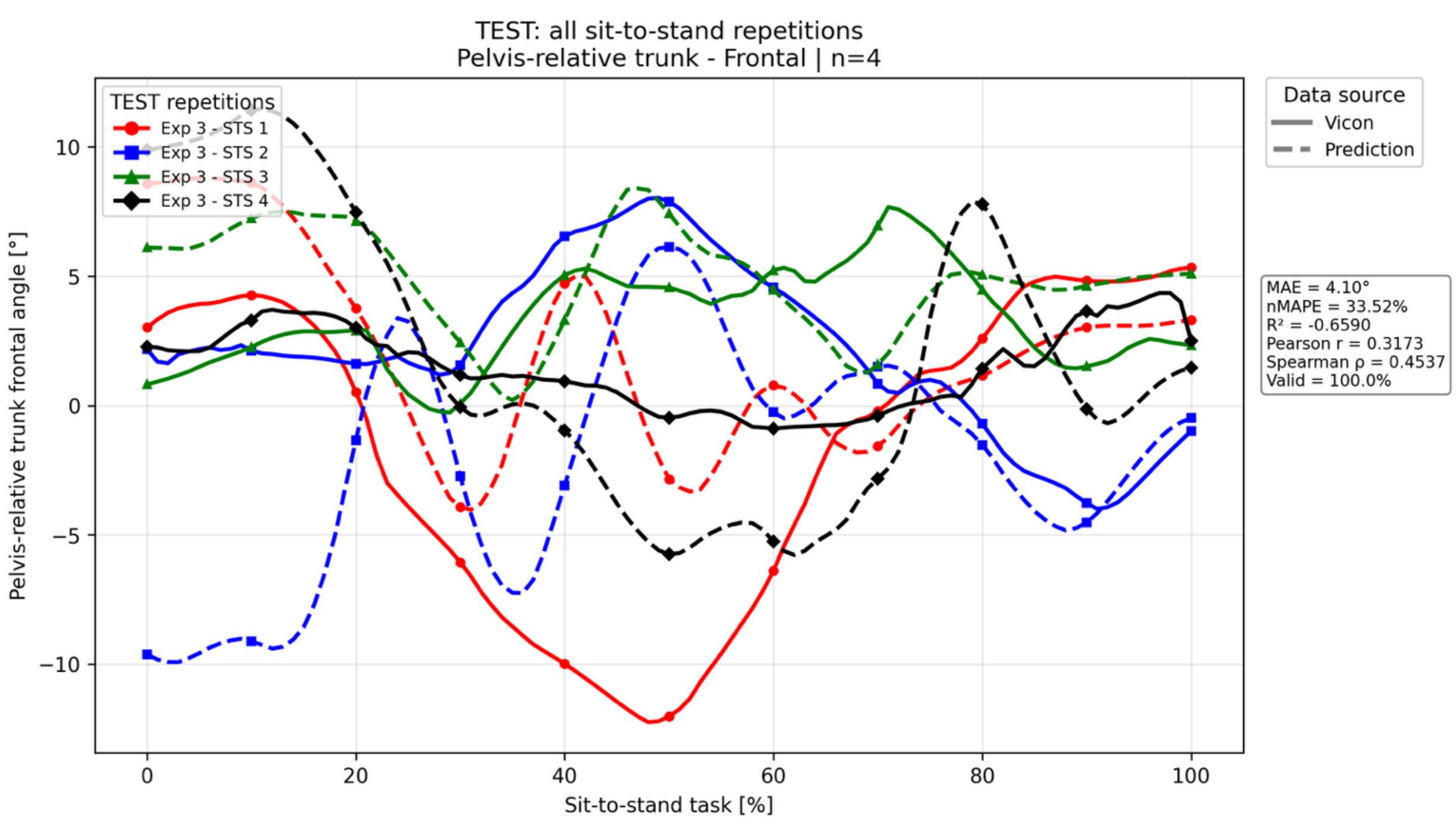

## Figure 4e | Projected knee descriptor, transverse

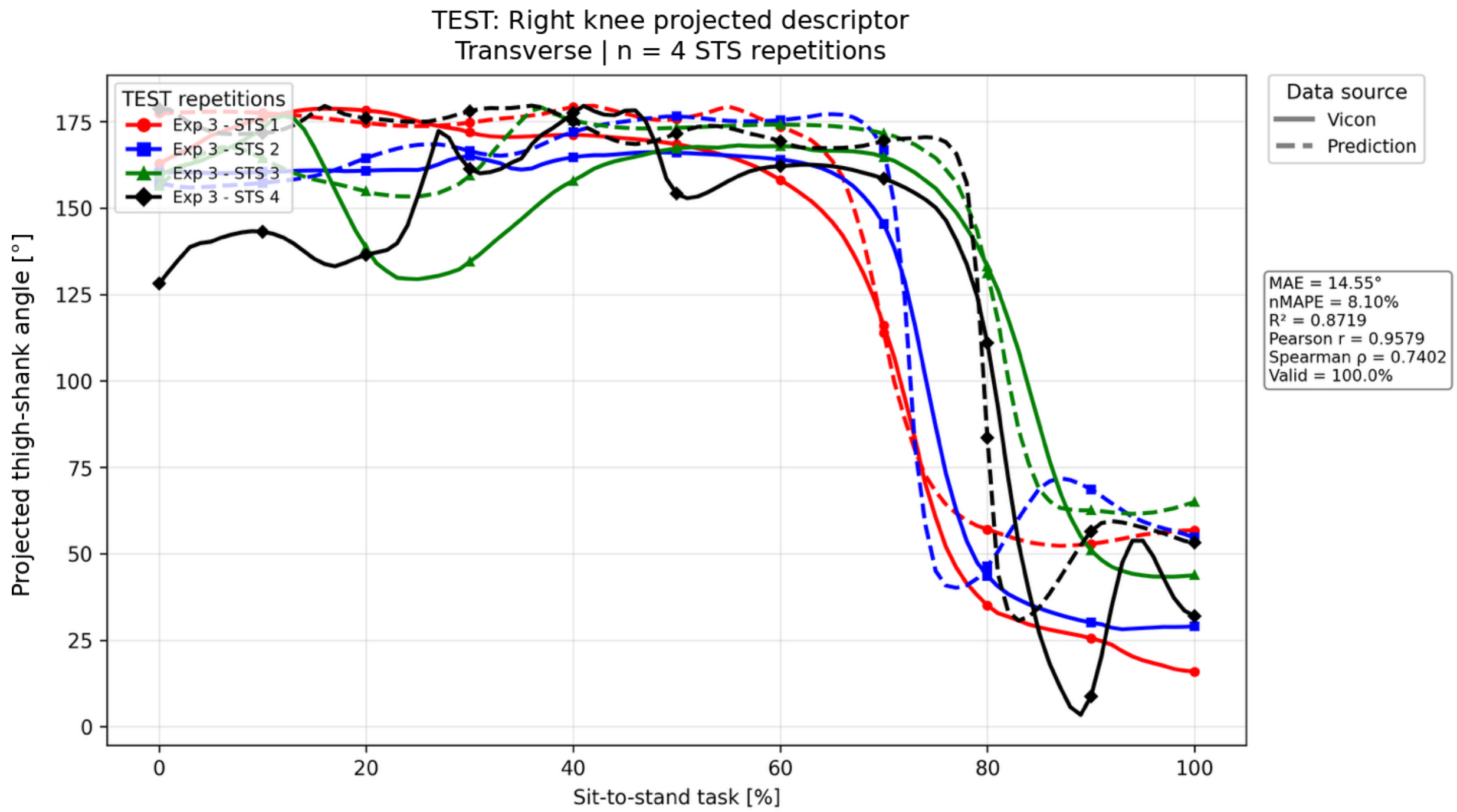

## Figure 4f | Trunk, transverse

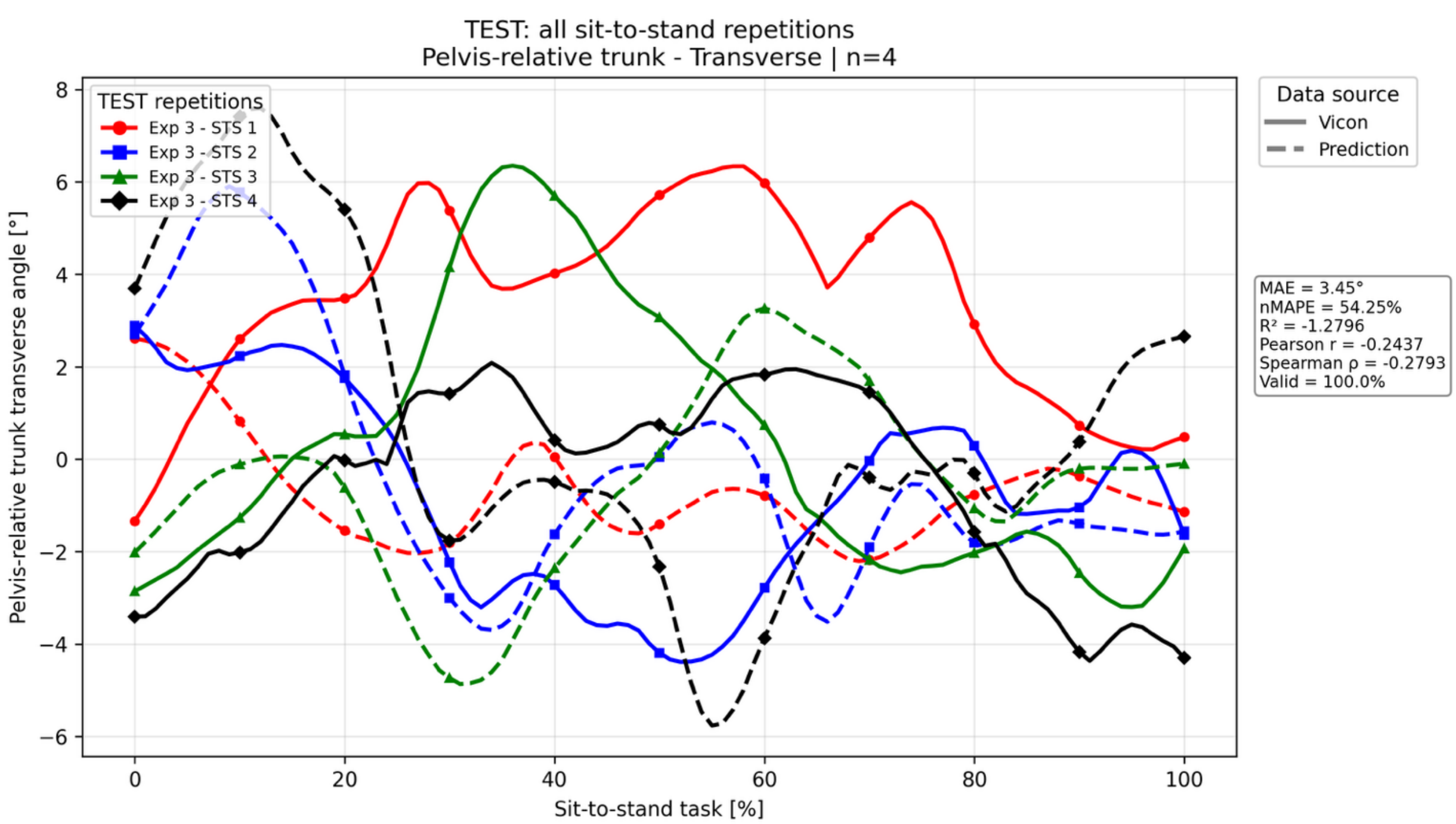

**Fig. 4 | Projected knee and pelvis-relative trunk descriptors during unassisted pediatric sit-to-stand.** a, c, e, Right-knee projected thigh-shank descriptors in the global sagittal, frontal, and transverse planes. b, d, f, Pelvis-relative trunk descriptors in the corresponding panels. Solid curves show the marker-based reference and dashed curves the predictions; matching colors identify the same repetition within each panel. All four test repetitions from the typically developing child are displayed. The knee quantities are projected geometric descriptors, not anatomical varus/valgus or internal/external rotation measurements. In c, the green reference curve (Exp. 3–STS 3) contains a transient excursion; in d, the red reference curve (Exp. 3–STS 1) contains a marked negative excursion. The processed predictions do not reproduce these excursions to the same extent. These intervals remain included in the reported agreement statistics. The plots alone do not establish the anatomical angle during the excursions or isolate the effect of rank reduction from post-processing. Numerical summaries are provided in the Results and Supplementary Table 3. n = 4 repetitions from one participant.

**Hip-knee cyclograms reveal differences not captured by pointwise error alone**

We next examined hip-knee cyclograms to determine whether reconstruction retained coupled-motion patterns in both pediatric cases (Fig. 5). In the child with cerebral palsy, hip and knee flexion-proxy mean absolute errors were 6.16° and 3.38° across four test repetitions, with a mean paired-angle discrepancy of 7.60°. Hip and knee $R^2$ values were 0.422 and 0.423, respectively. The reconstruction retained recognizable features of the coupled-motion pattern while showing differences in excursion and local trajectory geometry. Paired reference and predicted values were available for 80.4% of the normalized samples required for this cyclogram; the remaining 19.6% did not enter the numerical comparison. Prediction accuracy during reference-unavailable intervals could not be independently quantified against the marker-based system.

## Figure 5a | Cerebral palsy, right-sided proxies

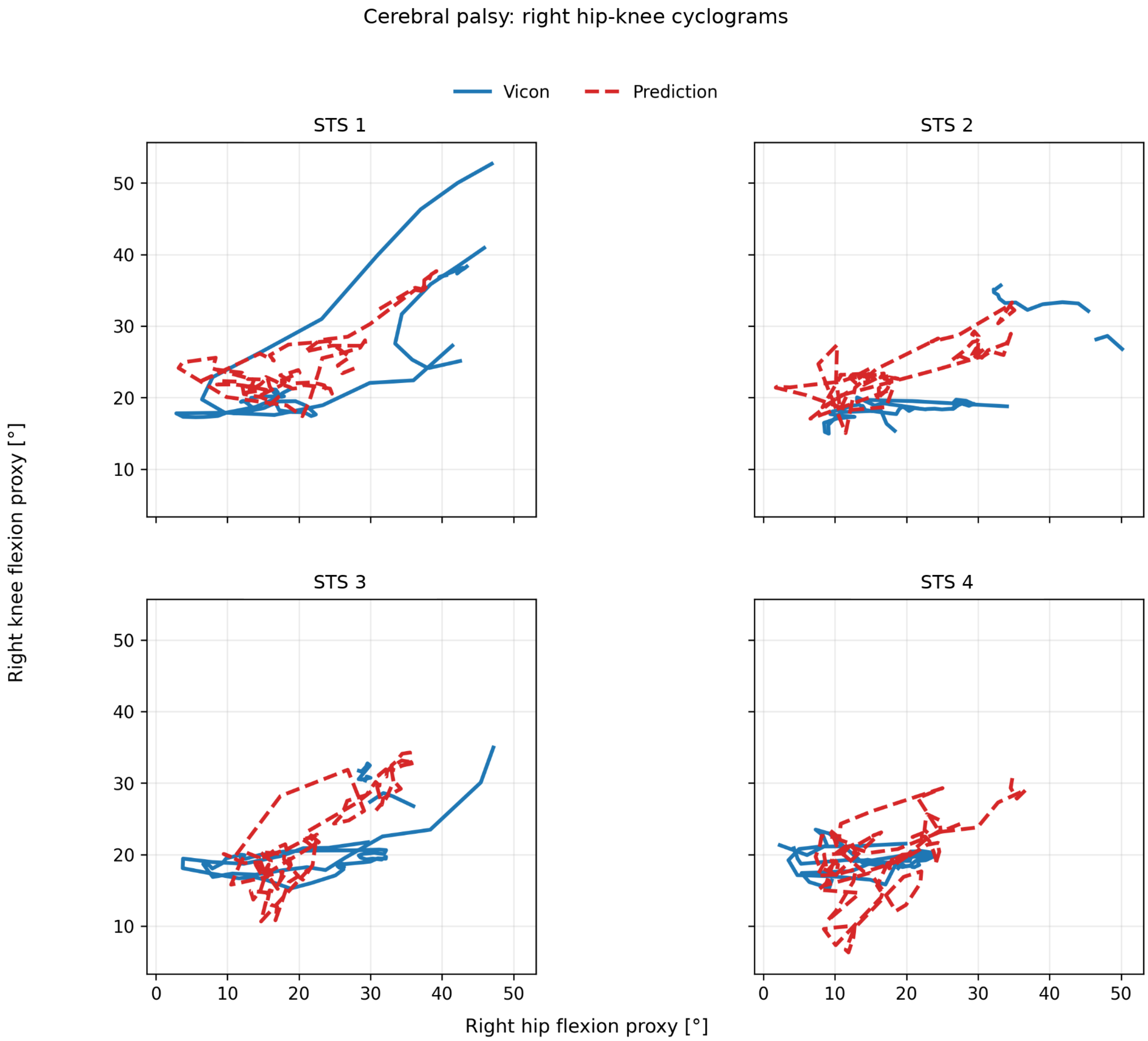

## Figure 5b | Typically developing, left-sided proxies

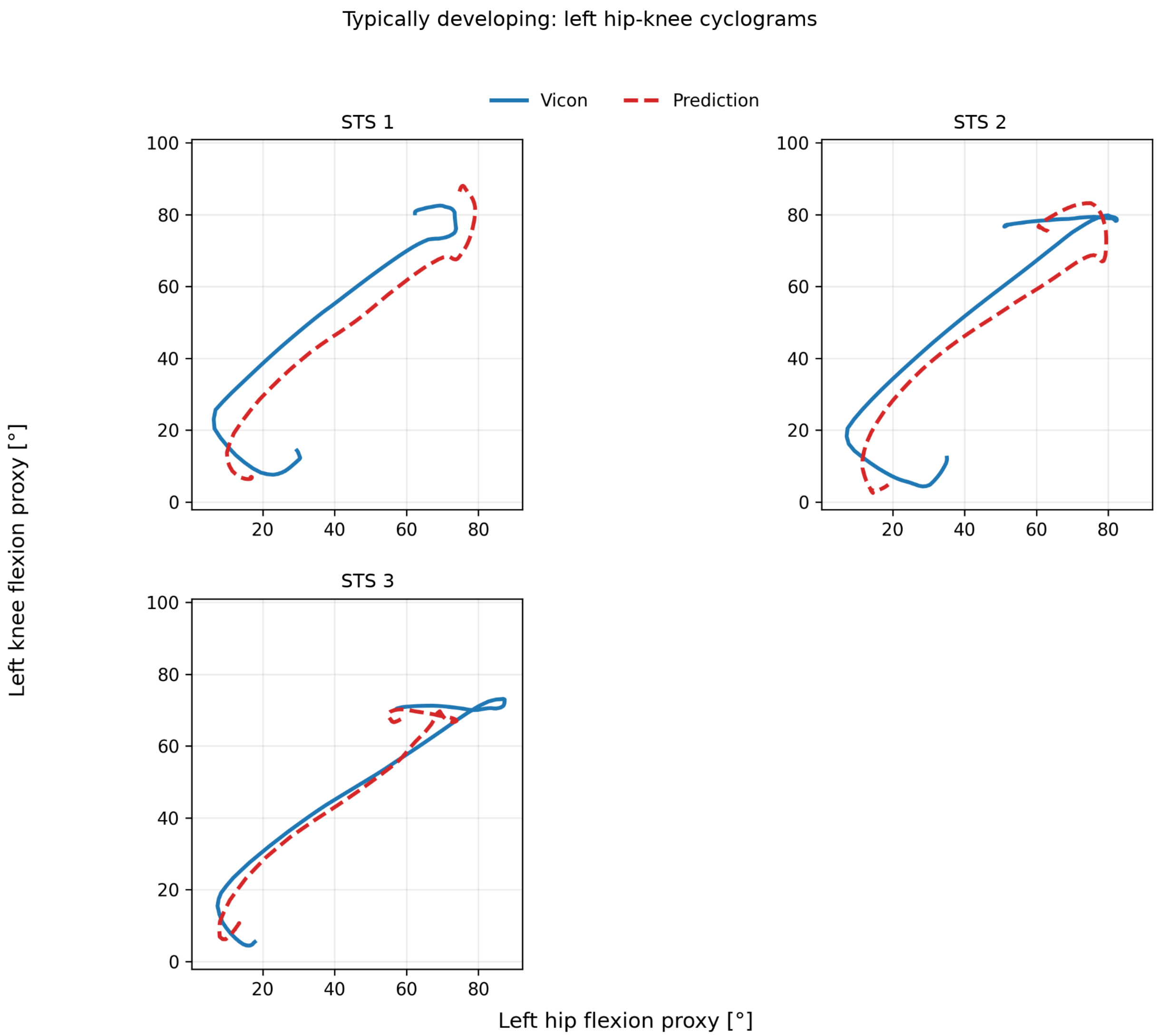

Typically developing: left hip-knee cyclograms
Vicon
Prediction
STS 1
STS 2
STS 3
Left knee flexion proxy [°]
Left hip flexion proxy [°]
0
20
40
60
80
100

**Fig. 5 | Individual hip-knee cyclograms in assisted and unassisted pediatric movement.** a, Right-sided hip and knee flexion-proxy cyclograms from four assisted test repetitions in the child with cerebral palsy (PC008). b, Left-sided cyclograms from three unassisted test repetitions in the typically developing child (TD005). Each repetition is displayed separately. Blue solid and red dashed curves indicate the marker-based reference and predicted paths, respectively; reference gaps are retained. The fourth position in b is intentionally empty, so no mean trajectory is counted as an additional trial. Axis limits are consistent within each case. The two cases are evaluated against their own reference trajectories, not as a same-limb or matched diagnostic-group comparison. Axes represent flexion proxies in degrees. Paired-data availability is 80.4% and 100% for the cerebral palsy and typically developing cases, respectively. Separate hip and knee errors, $R^2$ values, and paired-angle discrepancies are reported in the Results and Supplementary Table 3; the paired-angle discrepancy is not a positional error. n = 4 and n = 3 repetitions, respectively, each from one participant.

In the typically developing child, the cyclograms more closely reproduced the broad reference trajectory across three test repetitions. Hip and knee flexion-proxy mean absolute errors were 7.17° and 4.67°, with a mean paired-angle discrepancy of 9.41°. Hip and knee $R^2$ values were 0.879 and 0.960, respectively, with 100% paired-data availability. Together, these outcomes characterize complementary aspects of coupled-motion reconstruction: pointwise error and trajectory fit quantify numerical correspondence, while the cyclograms show the paired-angle paths.

### Multidirectional hopping reconstruction retains the overall knee-marker displacement profile

To examine a higher-speed and biomechanically distinct movement regime, the framework was also evaluated during the multidirectional single-leg hopping task. Toe-marker-derived flight intervals were identified for all training and test hops, with measures of agreement and prediction error observed in Table 1.

**Table 1 | Agreement between reference and predicted toe-marker-derived flight times.** Values are mean ± standard deviation. Bias is predicted minus reference flight time. MAE, mean

absolute error; RMSE, root mean square error. Each set comprises 16 hops from one participant. Since the video frequency was 50Hz, a single frame difference can generate a bias of 20 ms.

| Set | Valid Jumps | Marker-based reference flight time (s) | Predicted flight time (s) | Bias (s) | MAE (s) | RMSE (s) |
|---|---|---|---|---|---|---|
| TRAIN | 16/16 (100.0%) | 0.226 ± 0.030 | 0.240 ± 0.035 | 0.014 | 0.019 | 0.028 |
| TEST | 16/16 (100.0%) | 0.234 ± 0.044 | 0.242 ± 0.061 | 0.009 | 0.054 | 0.078 |

Across 16 test hops, the left-knee marker Z-coordinate trajectory had a mean absolute error of 20.96 mm, $R^2$ of 0.760, and Pearson correlation of 0.878 (Fig. 6). Together, these metrics indicate that the reconstruction reproduced the dominant rise-and-fall displacement pattern across held-out hops, with differences between reference and prediction in the amplitudes of individual repetitions. This finding extends the evaluation from pediatric transitional movements to a high-dynamic task and demonstrates preservation of a task-relevant positional trajectory across substantially different biomechanical demands.

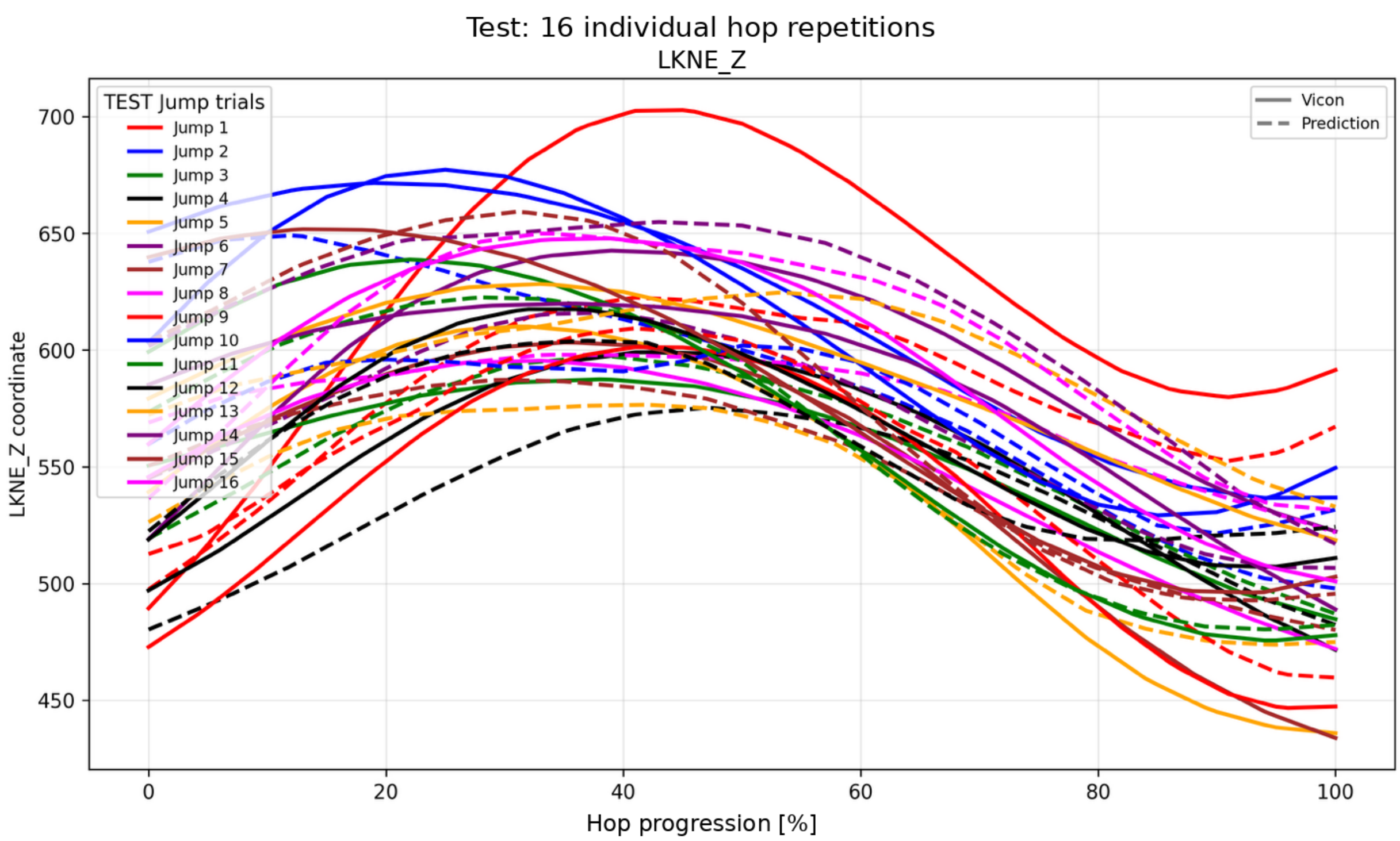

**Fig. 6 | Test reconstruction during multidirectional single-leg hopping.** Recorded Z-coordinate trajectories of the left-knee marker (LKNE) across 16 individual test hops. Solid

curves indicate the marker-based reference and dashed curves the predictions; colors and jump labels identify individual hops within the recorded sequence, not separate participants. All test hops are shown, including repetitions with amplitude discrepancies. The horizontal axis represents normalized hop progression. Coordinate-error summaries are provided in the Results and Supplementary Table 3. n = 16 hops from one adult participant.

**Discussion**

Quantitative three-dimensional kinematic assessment is fundamental to linking movement strategies to function in rehabilitation and to athletic performance; however, measurement demands become difficult to satisfy when movement includes atypical posture, physical assistance occluding markers, and intermittent loss of visibility.[1,4,6,7] We here evaluated a framework that combines rank-constrained latent representations with feedforward marker-coordinate regression to reconstruct movement from synchronized video.[18–20] Its contribution is the examination of reconstructed movement across positional, angular, and coupled-motion outcomes, connecting computational reconstruction with the biomechanical organization retained. Across test repetitions excluded from model training, the framework recovered dominant displacement profiles and sagittal-plane angular trajectories, while cyclograms revealed both preserved hip-knee relationships and limitations in reproducing individual coordination patterns. Together, these case-based findings support the feasibility of video-derived biomechanical reconstruction across clinically atypical, therapist-assisted pediatric movement, and high-speed multidirectional hopping, while demonstrating why recognizable movement progression and relatively small errors must be considered alongside preservation of coupled motion.

The assisted pediatric case addresses the practical challenge of examining functional movement while maintaining the support required to perform it. Sit-to-stand performance is associated with gross motor function and everyday mobility, and task-specific training can improve performance and selected self-care and mobility outcomes.[14,15,24] The relevance of retaining assistance during assessment is reflected in our stand-on power mobility trial protocol for children with cerebral palsy at GMFCS level III, which incorporates biomechanical assessment of functional tasks with hand support when independent performance is not possible.[25] Such protocols establish a need to characterize movement under the conditions in which children can participate. The present reconstruction of assisted sit-to-stand provides an initial basis for investigating that application, with physical support retained as part of the observed movement context. Its use for detecting rehabilitation-related change will require subsequent longitudinal evaluation.

The stronger sagittal-plane findings may partly reflect the larger joint and segment excursions associated with rising from sitting. These excursions may provide more distinguishable changes in the visual input and a broader movement signal for the model to learn than smaller frontal or transverse excursions. For descriptors with limited excursion, a modest angular discrepancy can represent a substantial proportion of the observed variation, helping explain why small absolute errors can coexist with weaker trajectory fit. Viewing geometry and the sensitivity of projected angular calculations may also contribute. These explanations remain hypotheses because their individual effects were not isolated. Related work incorporating activity-classification information into kinematic estimation demonstrates that movement-related constraints can improve reconstruction.[21] Here, the constraint acts on the learned representation rather than through explicit activity labels, and the present findings demonstrate the combined framework's ability to recover task-relevant movement features.

The cyclograms extend this interpretation by examining whether reconstructed angular quantities preserve their relationship throughout the transition. The typically developing case demonstrates that the framework can recover hip-knee coupling beyond reproducing individual trajectories. In the cerebral palsy case, smaller absolute paired-angle discrepancy coexisted with weaker trajectory fit and more pronounced shape differences. These findings describe different aspects of reconstruction: absolute discrepancy reflects distances between paired values, whereas cyclogram shape reflects how the two angular quantities evolve together. Smaller excursions could reduce absolute discrepancy without improving preservation of coupling, although this contribution requires examination of the corresponding movement ranges. Our prior pediatric sit-to-stand work similarly showed that recognizable cyclograms could coexist with incomplete recovery of individual patterns when movement extended beyond the principal observation plane.[6] The present findings therefore retain an important boundary: reconstruction of the overall transition does not, by itself, establish preservation of the individual coordination strategy.

The distinction presented above matters because coordination patterns in cerebral palsy depend on the movement demands being examined. During walking, Dussault-Picard and colleagues reported more in-phase knee-hip coordination during selected gait phases and associations between coordination measures and gait deviations.[26] Their subsequent uneven-surface study identified task-dependent changes in knee-hip coordination.[27] Both studies included children at GMFCS levels I and II, leaving direct extension to assisted transitions at level III unresolved.[26,27] These findings support examining individual coupling patterns within their task context; they do not establish that discrepancies between reconstructed and reference cyclograms are inherent to cerebral palsy. In the present cases, assistance, posture, movement excursion, and available training samples also differed. The scientific value of the cyclograms consequently lies in revealing which relationships are retained and where reconstruction remains incomplete.

The cyclograms observed for the child with cerebral palsy further illustrate the framework's capacity to reconstruct coupled movement despite interruptions in the marker-derived record. In STS 2 (Fig. 5a, upper right), the processed video-derived prediction maintained a continuous hip-knee cyclogram throughout the transition, including the interval in which the marker-derived trajectory was discontinuous. This finding extends the demonstrated capability beyond individual coordinate prediction to the reconstruction of a continuous relationship between angular quantities during assisted functional movement. The transient marker-derived excursions in the frontal-plane descriptors (Fig. 4c,d), which were not reproduced to the same extent by the computational approach, likewise highlight the value of evaluating reconstructed movement alongside the quality of the available reference. Together, these observations support the potential of the complete framework to retain interpretable movement representations when marker-based measurements are compromised. The accuracy of reconstructed features within reference-unavailable intervals remains to be independently established, with temporal smoothing considered part of the demonstrated pipeline.

The hopping demonstration extends the evaluation to repeated propulsion, flight, landing, and redirection. Recovery of the dominant knee-marker displacement pattern across test repetitions supports reconstruction of positional progression during a movement sequence with substantially different demands from sit-to-stand. Remaining amplitude differences indicate that reproducing the sequence does not ensure equivalent recovery of its excursion. This demonstration provides a basis for further investigation of video-derived trajectories during rapid, repeated movement; the selected positional outcome does not establish reconstruction of whole-body landing mechanics or athletic performance. Moreover, application to separately trained task datasets demonstrates the use of the framework under distinct conditions rather than direct transfer between tasks. The lower training errors for the back-marker trajectories, pediatric cyclograms, and hopping outcome reinforce test-repetition performance as the appropriate basis for interpreting these demonstrations (Supplementary Figs. 1, 4–6).

Several considerations define the scope of this proof of concept. Each displayed test condition was represented by one participant, allowing examination of repeated movements within these cases rather than between-participant generalizability. The evaluated pipeline included multiple views, rank-constrained representation learning, marker-coordinate regression, and post hoc temporal smoothing; the contribution of each component remains unresolved. Natural occlusion was present but was not quantified by duration, anatomical distribution, or camera view. For the cerebral palsy cyclogram, paired values were available for 80.4% of normalized samples; this outcome-specific availability does not characterize occlusion across the dataset or establish reconstruction accuracy during reference-unavailable intervals. The two-camera configuration and video sampling at 50 Hz, compared with the 100 Hz marker-based reference, further define the acquisition conditions examined and the temporal detail available during rapid movement.

Interpretation also depends on the reference measurements and visual information present during acquisition. Marker placement, soft-tissue motion, and gap handling influence marker-derived trajectories, which provide an experimental reference rather than direct observation of skeletal motion.[1,4] Reflective markers and the assisting examiner were visible, and their potential contribution to the learned predictions requires separate investigation before marker-free performance or independence from examiner-related visual context can be established. The angular descriptors and flexion proxies likewise characterize reconstructed movement relationships; their anatomical joint-rotation validity was not established. These considerations are particularly important when extending a computational demonstration toward measurements intended to inform individual clinical interpretation.

The next step is participant-level evaluation under documented visibility and acquisition conditions, with simpler reconstruction models and component-level comparisons used to determine where rank-constrained representations add benefit. Analyses that consider angular excursion, cyclogram shape, and reference continuity together would further clarify why coupled-motion reconstruction differs across repetitions. Explicit modeling of latent temporal evolution offers a complementary direction beyond frame-wise prediction and subsequent filtering.[18] The present study provides a case-based foundation for this progression by demonstrating recovery of task-relevant movement features while identifying the additional demands of preserving their relationships. Its broader contribution is to make movement organization an explicit outcome of reconstruction, so that computational performance is judged through both the trajectories recovered and the biomechanical information they retain.

## Methods

### Ethical oversight and participant selection

Human subjects data collection was conducted in accordance with ethical regulations for human subjects research and was approved by the institutional review board at the University of North Florida. Participants included two children with cerebral palsy (males, 4 and 5 years old; Gross Motor Function Classification System level III) under protocol LEG-2147539, two typically developing children (males, 4 and 8 years old) under protocol FY2024-33, and one female young adult (24 years old) under protocol LEG-1957720. The cases were selected to examine assisted pediatric movement, an unassisted pediatric comparator, and a high-dynamic movement task rather than to estimate population-level performance. Written informed consent was obtained from the adult participant and legal guardians of the children, with child assent obtained as required by the approved protocols.

The reported pediatric participants and recordings analyzed in the present study were distinct from those included in our previously published multimodal motion-capture work.[6] Although the acquisition procedures followed the same general experimental framework, no participant or trial-level overlap occurred between the two datasets.

### Task-specific data acquisition

For sit-to-stand, children performed consecutive cycles as quickly as possible from a height-adjustable chair without armrests, with seat height standardized to approximately 90° hip and knee flexion. The typically developing child performed the task with hands on hips. For the child with cerebral palsy, an examiner positioned in front provided hand-held assistance to support task completion and additional hip support when needed for safety. Assistance was part of the task condition and was not assumed to be mechanically neutral. The analysis examined movement performed with that support rather than an estimate of how the child would move without assistance.

The multidirectional hopping task (CUBE) was performed on a 3 by 3 cloth grid composed of 0.45 m squares. The participant hopped on one leg from the center square to each surrounding square, beginning with the front square and proceeding clockwise, returning to the center after each outward hop. Instructions were to complete the sequence as quickly as possible without losing balance. Three sequences were collected on the right leg and three on the left. A complete sequence contains 16 individual hops. The reported reconstruction plots contain 16 training hops and 16 test hops; the displayed training and test recordings are identified in Supplementary Table 1.

**Motion capture and synchronization**

Laboratory configuration followed the previously published multimodal acquisition protocol.[6] Marker trajectories were recorded using twelve Vicon Vero-series infrared cameras at 100 Hz. Synchronized video was recorded at 50 Hz using two FLIR Blackfly cameras positioned to provide complementary views while accommodating examiner assistance. Hardware synchronization used the Vicon LockLab unit to deliver a transistor-transistor logic trigger to both systems at the start of each trial. Modalities were aligned to a common time base for comparison. Camera geometry, calibration, resampling, and time-normalization details are specified in Supplementary Methods.

**Reference trajectories and kinematic descriptors**

Reflective markers were placed by an experienced clinician according to the Vicon Plug-in Gait full-body conventions.[2,3,28] The positional outcomes presented here are the recorded right-back marker (RBAK) X, Y, and Z coordinates and the left-knee marker (LKNE) Z coordinate. These are marker trajectories, not anatomical joint-center estimates. Coordinate errors are reported in millimeters, and X, Y, and Z retain the acquisition coordinate labels rather than re-labelled as anatomical planes.

Angular outcomes comprised marker-derived right-knee and pelvis-relative trunk descriptors in three planes, together with the hip and knee flexion proxies used to construct the cyclograms. The knee descriptors were calculated as the unsigned included angle between the thigh and shank vectors after projection onto the global sagittal, frontal, and transverse planes. Accordingly, these quantities are reported as projected inter-segment angle descriptors and are not interpreted as anatomical varus/valgus or internal/external rotation angles. The hip and knee flexion proxies used in the cyclograms were likewise retained as marker-derived geometric descriptors rather than joint-coordinate-system rotations. Each predicted descriptor was compared with the corresponding quantity calculated from the marker-based reference trajectories using the same code and geometric formulation. The geometric definition, reference frame, sign convention, missing-data treatment, and distinction from native Plug-in Gait outputs are provided in the Supplementary Methods.

**Video preprocessing and computational framework**

The two video streams were aligned using the synchronization trigger and trimmed to task-specific analysis windows. Sit-to-stand windows included initiation, ascent, and stabilization; hopping windows included take-off, flight, and landing. Each view was cropped to a region encompassing the participant and, when present, the assisting examiner. Motion-sensitive channels generated through background subtraction supplemented the RGB frames as multichannel inputs. Video preprocessing, including frame handling, region-of-interest cropping, and background subtraction, was implemented in Python using OpenCV[29]. The specific background-subtraction procedure, crop dimensions, scaling, and software versions are provided in Supplementary Methods.

The reconstruction framework combined view-specific convolutional Rank-Reduction autoencoders with a feedforward multilayer perceptron (MLP). The autoencoders learned nonlinear latent representations whose effective rank was constrained through truncated singular value decomposition.[18–20] The combined model mapped the video-derived representations to target marker coordinates. Frame-wise predictions were followed by post hoc temporal smoothing, and the reported outcomes characterize the complete reconstruction and post-processing pipeline. The encoder and decoder structures, latent-rank selection, regression network, loss function, optimization procedure, and inference specification are described in Supplementary Methods and Supplementary Fig. 8.

Within the rank-reduction component, the encoder maps the input representation to a latent matrix, and truncated singular value decomposition is applied during training to retain a lower-rank latent approximation.[18,19] The effective latent dimensionality is therefore defined by the retained rank rather than solely by a predefined dimensionality of the latent layer. Following training, the RRAE inference procedure fixes a basis derived from the training data and projects newly encoded observations onto that basis.[18,19] Each available video view was supplied to a separate convolutional RRAE. The RRAEs were pretrained, and their rank-constrained representations were concatenated into a single vector supplied to the MLP to predict marker locations in X, Y, and Z coordinates. Input pixel values were normalized to [0, 1], and target coordinates were scaled to [−1, 1] using min-max scaling. Training combined mean squared errors for predicted marker coordinates and reconstructed pixel values with a perceptual loss using pretrained VGG19 features[30] and L2 regularization of network weights and biases. The RRAEs and MLP were then jointly fine-tuned for up to 2,000 epochs using the Adam optimizer with early stopping. Kinematic descriptors were calculated from the predicted marker coordinates after training using the geometric relationships specified for each outcome.

**Data partitioning and scope of evaluation**

The reported computational datasets contained 8, 4, and 8 recordings for the cerebral palsy, typically developing, and hopping conditions, respectively. For each condition, one recording was held out for testing. Within the remaining recordings, the first 80% of the temporal sequence was used for training and the remaining 20% for validation. Frames lacking required reference measurements were used for prediction but were not assigned supervised agreement metrics. The held-out recording was used for model evaluation after tuning. Separate condition-specific models were constructed for the cerebral palsy, typically developing, and hopping datasets.

The test figures contain four assisted repetitions from the child with cerebral palsy, four typically developing repetitions for the knee and trunk descriptors, three typically developing repetitions for the cyclogram, and 16 individual hops from the adult case. Training figures contain five assisted repetitions, four unassisted repetitions, and 16 individual hops. Analysis-specific recording labels and repetition counts are summarized in Supplementary Table 1. No participant-level or cross-task generalization is inferred from these counts.

### Error metrics and descriptive analysis

Mean absolute error was the primary reported measure of coordinate- or descriptor-specific discrepancy. $R^2$ and Pearson correlation described complementary aspects of trajectory fit and association; they were not treated as interchangeable measures of agreement. Errors were kept separate by outcome and unit rather than pooled across positions and angles. Hip-knee cyclograms paired the two flexion proxies, with hip and knee mean absolute errors and the mean paired-angle discrepancy reported in degrees. This latter quantity is a distance in the two-angle representation, not a positional error in millimeters. Exact calculations, aggregation across repetitions, and handling of invalid samples are specified in Supplementary Methods. Only held-out test-recording results are reported in Figures 3 to 6.

Results are descriptive. Repetitions and time-normalized samples remain nested within recordings and participants. The available dataset comprises five participants; the displayed test outcomes represent three individual cases. The main figures display individual test trajectories to make within-case variation visible, and training trajectories are presented separately in the Supplementary Information. No population-level hypothesis testing was performed in the reported analyses. Paired-data availability was 80.4% for the cerebral palsy test cyclogram and 100% for the typically developing test cyclogram. This indicator describes the normalized samples entering each paired comparison; it is not a global measure of body or video-view occlusion.

### Data availability

The de-identified data supporting the findings of this study are not publicly available because the informed consent documents and institutional approvals did not include unrestricted public data sharing. The data may be made available by the corresponding author upon reasonable request, subject to institutional review, applicable ethical requirements, and execution of an appropriate data-use agreement.

### Code availability

The custom code used for the computational analyses will be made publicly available in a permanent repository upon publication. The code is available from the corresponding author upon reasonable request during peer review.

**Acknowledgements and funding**

Data collection for the cerebral palsy component was supported by the Eunice Kennedy Shriver National Institute of Child Health and Human Development of the National Institutes of Health under award R03HD114548 and by the Foundation for Physical Therapy Research under project 1189205, both awarded to G.M.C. No dedicated funding was received for the computational development and secondary analyses reported here. The content is solely the responsibility of the authors and does not necessarily represent the official views of the National Institutes of Health.

**Author contributions**

Conceptualization, C.G. and G.M.C.; methodology, C.G., L.Y.M. and G.M.C.; software, C.G. and L.Y.M.; validation, C.G., L.Y.M. and G.M.C.; formal analysis, C.G. and L.Y.M.; investigation, L.Y.M. and G.M.C.; resources, C.G., L.Y.M. and G.M.C.; data curation, L.Y.M.; original draft preparation (writing), C.G., L.Y.M. and G.M.C.; review and editing (writing), C.G., L.Y.M. and G.M.C.; visualization, C.G.; supervision, G.M.C.

**Competing interests**

The authors declare no competing interests.

**Supplementary Information**

# Human movement reconstruction through latent structural representations during assisted transitions and multidirectional hopping

## Supplementary Note 1 | Organization of recordings and analyzed repetitions

The analyses are organized by participant, acquisition recording, and movement repetition. Counts identify the repetitions represented in the reconstruction figures rather than the number of independent biological samples. The displayed pediatric cases contribute two participants, the data of one of them is used only for training and the other one is the one reported in the manuscript as is included a combination of train and test sets. The adult hopping case contributes one participant; the complete available dataset is described in the main Methods. Distinguishing complete acquisition trials from repetitions is particularly important for the CUBE sequence, because each outward hop and each return to the center is a separate movement within the sequence.

**Supplementary 1 | Analysis-specific recording and repetition counts for the reported participants.**

| Case and outcome | Training | Test | Recording labels and analysis scope |
|---|---|---|---|
| Cerebral palsy: RBAK coordinates | 5 STS repetitions | 4 STS repetitions | Training examples: PC008_baseline_STS02 / STS03. Test: PC008_baseline_STS01. |
| Cerebral palsy: right hip-knee proxies | 5 STS repetitions | 4 STS repetitions | Training aggregate labels: Exp 5 and Exp 6. Test aggregate label: Exp 4. |
| Typically developing: knee and trunk descriptors | 4 STS repetitions | 4 STS repetitions | Training: TD005_sts02. Test: TD005_sts03. Knee aggregate training panels are labelled left; test panels are labelled right. |
| Typically developing: left hip-knee proxies | 4 STS repetitions | 3 STS repetitions | Training: TD005_sts02. Test: TD005_sts03. |

| Case and outcome | Training | Test | Recording labels and analysis scope |
|---|---|---|---|
| Adult: LKNE Z coordinate | 16 individual hops | 16 individual hops | Training examples: ACL001_cube01. Test examples: ACL001_cube02. |

**Supplementary Note 2 | Training reconstruction**

Training trajectories are presented separately from the test evaluation because fit to data used in model development does not establish performance on new recordings. For the cerebral palsy back-marker coordinates, training mean absolute errors were 2.67, 6.40, and 3.13 mm for X, Y, and Z, respectively. The corresponding test values were larger for all three components. Training hip-knee proxy errors were also lower than the test summaries in each pediatric case, although the cerebral palsy training cyclogram had a lower paired-data availability. In the hopping case, training LKNE Z-coordinate mean absolute error was 5.98 mm, compared with 20.96 mm in testing. These observations describe training-to-test differences in the reported analyses and are not evidence of participant-level generalization.

**Supplementary Table 2 | Training-set error and trajectory-fit summaries.**

| Outcome | MAE | $R^2$ | Additional information |
|---|---|---|---|
| CP RBAK X | 2.673 | 0.9866 | 5 repetitions; mm |
| CP RBAK Y | 6.395 | 0.9695 | 5 repetitions; mm |
| CP RBAK Z | 3.127 | 0.9838 | 5 repetitions; mm |
| TD knee, frontal | 1.14° | 0.8157 | 4 repetitions; aggregate labelled left |
| TD knee, sagittal | 1.66° | 0.9945 | 4 repetitions; aggregate labelled left |
| TD knee, transverse | 3.69° | 0.9787 | 4 repetitions; aggregate labelled left |
| TD trunk, frontal | 1.60° | 0.9332 | 4 repetitions; pelvis-relative |
| TD trunk, sagittal | 1.96° | 0.9676 | 4 repetitions; pelvis-relative |
| TD trunk, transverse | 1.05° | 0.7065 | 4 repetitions; pelvis-relative |

| Outcome | MAE | $R^2$ | Additional information |
|---|---|---|---|
| CP right hip proxy | 0.53° | 0.989 | 5 repetitions; pair discrepancy 0.86°; availability 62.4% |
| CP right knee proxy | 0.73° | 0.860 | Same cyclogram analysis |
| TD left hip proxy | 1.96° | 0.992 | 4 repetitions; pair discrepancy 2.67°; availability 100% |
| TD left knee proxy | 1.59° | 0.995 | Same cyclogram analysis |
| Adult LKNE Z | 5.98 | 0.982 | 16 hops; mm |

CP, cerebral palsy; TD, typically developing; MAE, mean absolute error. The knee training aggregate panels and test panels are labelled with different sides, so those values are not interpreted as a matched limb-level training-to-test comparison. Coordinate errors are reported in millimeters. Angular values refer to the plotted descriptors or proxies.

**Supplementary Note 3 | Outcome-specific test performance**

Test outcomes are retained separately to avoid averaging over fundamentally different units and movement descriptors. The positional analyses concern specific physical-marker coordinates, whereas angular and cyclogram analyses concern derived kinematic quantities. Pearson correlation and the reported $R^2$ are shown separately where available. The difference is relevant for outcomes such as CP RBAK Y, for which positive correlation coexists with a negative $R^2$. Errors are reported separately for each outcome.

**Supplementary Table 3 | Test-set error, fit, and paired-data availability.**

| Outcome | MAE | $R^2$ | Pearson r or availability |
|---|---|---|---|
| CP RBAK X | 26.999 | 0.5636 | r = 0.8865; n = 4 |
| CP RBAK Y | 59.468 | -0.1639 | r = 0.8130; n = 4 |
| CP RBAK Z | 17.758 | 0.8350 | r = 0.9188; n = 4 |
| TD right knee, frontal | 4.46° | 0.4252 | r = 0.7105; availability 100%; n = 4 |

| Outcome | MAE | $R^2$ | Pearson r or availability |
|---|---|---|---|
| TD right knee, sagittal | 5.65° | 0.9441 | r = 0.9781; availability 100%; n = 4 |
| TD right knee, transverse | 14.55° | 0.8719 | r = 0.9579; availability 100%; n = 4 |
| TD trunk, frontal | 4.10° | -0.6590 | r = 0.3173; availability 100%; n = 4 |
| TD trunk, sagittal | 5.38° | 0.8175 | r = 0.9197; availability 100%; n = 4 |
| TD trunk, transverse | 3.45° | -1.2796 | r = -0.2437; availability 100%; n = 4 |
| CP right hip proxy | 6.16° | 0.422 | Pair discrepancy 7.60°; availability 80.4%; n = 4 |
| CP right knee proxy | 3.38° | 0.423 | Same cyclogram analysis |
| TD left hip proxy | 7.17° | 0.879 | Pair discrepancy 9.41°; availability 100%; n = 3 |
| TD left knee proxy | 4.67° | 0.960 | Same cyclogram analysis |
| Adult LKNE Z | 20.96 | 0.760 | r = 0.8781; n = 16 individual hops |

Counts refer to movement repetitions within one displayed participant per condition. Coordinate errors are in millimeters; angular errors and paired-angle discrepancies are in degrees. Paired-data availability describes the finite reference-prediction samples included in each outcome, not global body or video-view occlusion. A shared descriptor definition and inclusion mask were used for each paired comparison.

## Supplementary Methods

### Acquisition, calibration, and temporal alignment

Data were acquired using the synchronized Vicon system and two-camera video configuration described in the main Methods and based on the published multimodal protocol.[1] The Vicon sampling rate was 100 Hz and the video sampling rate was 50 Hz, with a shared trigger at trial onset. Repetitions were aligned to normalized movement progression for comparison.

### Reference processing and geometric definitions

Positional outcomes were evaluated at the skin-marker locations identified as RBAK and LKNE in the Plug-in Gait marker conventions.[2] The hip and knee flexion proxies are reported separately from positional trajectories and from the plane-labelled knee descriptors.

Recordings containing missing marker measurements were retained, with the corresponding reference-unavailable time steps treated as prediction-only observations for supervised agreement evaluation. Input data were not temporally filtered. Predicted outputs were temporally smoothed after reconstruction using Savitsky-Golay filter with a window length of 10 time-steps and a polynomial order of 2.

The knee descriptors were calculated as unsigned included angles between the thigh and shank vectors after projection onto the global sagittal, frontal, and transverse planes. These quantities are projected inter-segment descriptors rather than anatomical joint-coordinate-system rotations. Reference and predicted coordinates were processed through the same custom geometric calculation. Checks for nearly degenerate projected vectors did not lead to sample exclusion in the reported analyses. The signed pelvis-relative trunk descriptors and the hip and knee flexion proxies were evaluated as separate derived quantities.

### Video representation and model implementation

RGB images were supplemented by motion-sensitive channels generated by background subtraction. The representation was used in a framework combining convolutional rank-reduction autoencoding and feedforward marker-coordinate regression. Published RRAE formulations provide the methodological basis for rank-constrained latent representations,[3–5] whereas the architecture and training configuration for the present video-to-coordinate mapping are study-specific.

Each original RGB view was supplemented by a background-removed RGB view retaining moving pixels. The two three-channel images were concatenated along the channel dimension to form a six-channel input tensor for the corresponding RRAE, by order R-G-B channels for the full image followed by the R-G-B channels for the background-removed images.

The RRAE formulation encodes the input tensor into a latent matrix and uses truncated singular value decomposition during training to retain a rank-constrained latent representation.[3,4]

During inference, newly encoded observations are projected onto a basis identified from the training data rather than subjected to an SVD fitted to each test sample.[3,4] The view-specific representations are concatenated and supplied to the coordinate regressor. The deterministic representation used here is distinct from the variational sampling and explicit latent-dynamics extensions described in the cited methodological work.

**Supplementary Table 4 | Implementation information for reproducibility.**

| Component | Study-specific information |
|---|---|
| Video inputs | Two synchronized views were represented by original RGB and foreground RGB channels (Three views for the cerebral palsy use case). Cropping retained the region of interest in each view, and images were spatially down sampled to reduce memory requirements. Each view had six channels. |
| Encoder / decoder | The supplied cerebral palsy architecture is shown in Supplementary Fig.8: a, convolutional encoder; b, image decoder; c, feedforward coordinate regressor. The illustrated encoder has a 10,000-entry latent layer; retained rank is reported separately below. Figure 8(c) illustrates the CUBE use case, while the feedforward network changes only its input and output layer dimensions for the typically developing kids use case to 20,000 and 63 respectively, and to 30,000 and 63 respectively for the Cerebral palsy use case. |
| Latent representation | The reported configuration retained a rank of 15 SVD modes with a batch size of 30. Retained rank was selected through training and validation comparisons. The latent representation used two videos encoded and their latent spaces concatenated for the CUBE and typically developing children use cases, while Cerebral Palsy use case used three videos for three views, with latent spaces concatenated. |
| Optimization | During the final training, the VGG19 perceptual loss is weighted by a value of $10^{-6}$, the reconstructed pixel values are weighted by $10^{-2}$, and the marker-coordinate loss by a factor of 1. The optimizer was Adam, implemented with a custom learning rate loop adapting the learning rate from $10^{-3}$ to $10^{-6}$ during the training. The batch size was 30, with a maximum of 2,000 epochs. Early stopping used a patience of 10 epochs without validation-loss improvement. The VGG19 feature extractor follows Simonyan and Zisserman,[6] the frozen pretrained VGG19 selected the output of layer "block4_conv3" for the perception loss. |

| Component | Study-specific information |
| --- | --- |
| Data partitions | One recording per condition was held out for testing. Within the remaining recordings, the first 80% of each temporal sequence was used for training and the last 20% for validation. Separate models were fitted to the cerebral palsy, typically developing and hopping (CUBE) datasets. |
| Inference | Frame-wise coordinate predictions were followed by temporal post-processing using Savitsky-Golay filtering with a window size of 10 and polynomial order of 2, before the reported biomechanical evaluation. |
| Execution and release | The reported implementation used Python 3.11.9 and TensorFlow 2.16.2. |

**Error aggregation and paired-data availability**

Prediction accuracy was assessed using MAE, $R^2$, Pearson's $r$ correlation, and Spearman's $\rho$, computed from paired finite reference and predicted samples. Within-case metrics were calculated from pooled valid normalized samples across repetitions rather than from mean curves or averages of repetition-level metrics; therefore, repetitions contributed in proportion to their number of valid samples. No correction for temporal autocorrelation was applied, so these statistics were treated as descriptive agreement measures.

For hip-knee cyclograms, a normalized sample was included only when the reference and predicted hip and knee flexion-proxy values were all finite. Paired-angle discrepancy was the mean Euclidean distance between reference and predicted hip-knee points over those paired samples. Paired-data availability was the percentage of included normalized samples relative to all normalized samples across the analyzed repetitions. Sit-to-stand trajectories were normalized over the full 0 to 100% interval, with comparisons restricted to contiguous finite segments and no interpolation across missing-data gaps. Missing marker measurements propagated to nonfinite reference descriptors; no prediction-confidence threshold was used. This availability measure is outcome-specific and is not a global measure of body or video-view occlusion but for the specific considered markers for the task at hand.

Individual repetitions are displayed separately to show the variability and reconstruction discrepancies within each case. Results are descriptive; repeated movements within a participant do not supply independent biological replicates.

## Supplementary figures

**Supplementary Fig. 1 | Training right-back marker reconstruction in the cerebral palsy case.**

a–c, X, Y, and Z RBAK coordinates for five training sit-to-stand repetitions. Solid curves show the marker-based reference and dashed curves the predictions; colors distinguish repetitions. Training fit is reported separately from the test reconstruction in main Fig. 3. n = 5 repetitions from one participant.

**Supplementary Figure 1a**

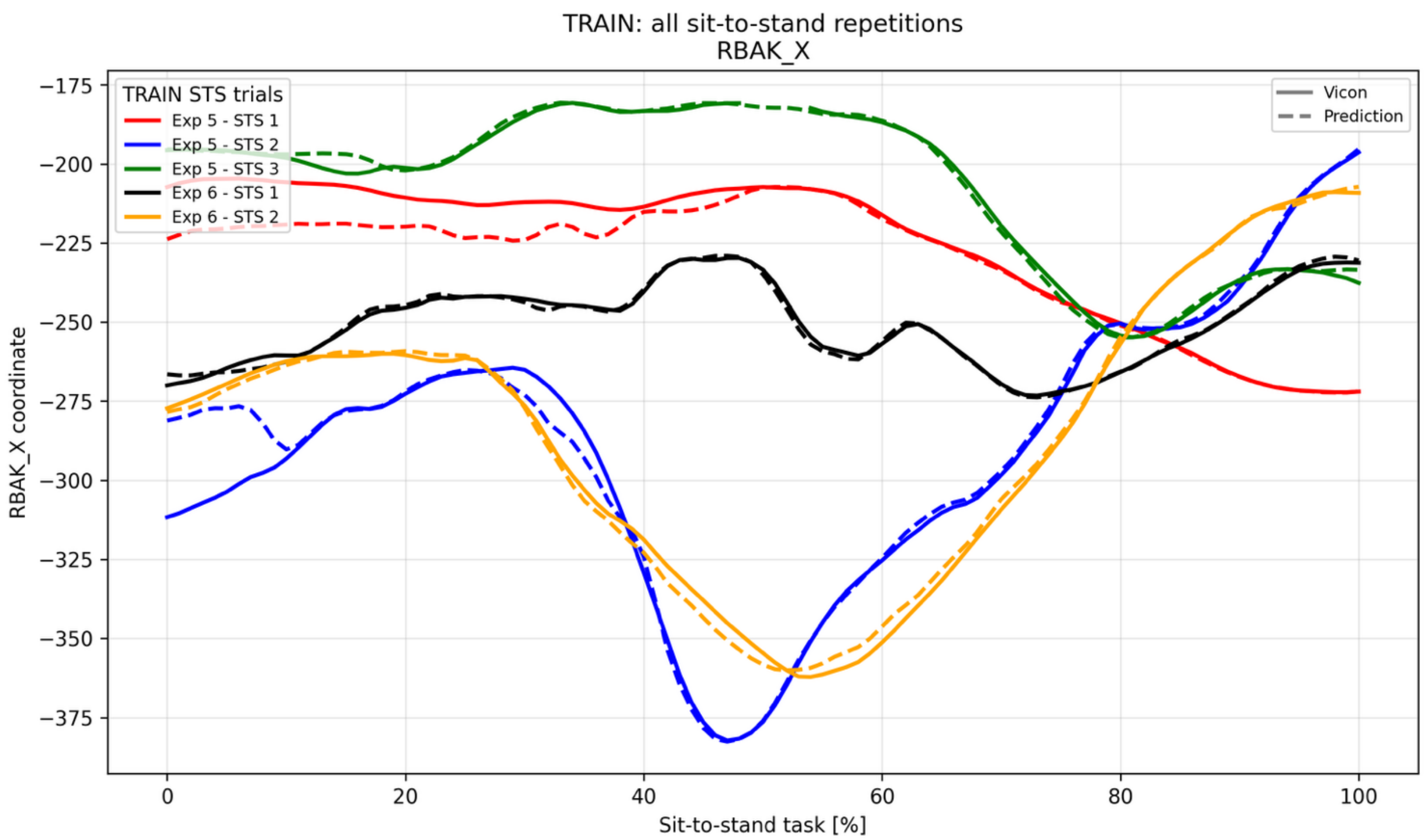

## Supplementary Figure 1b

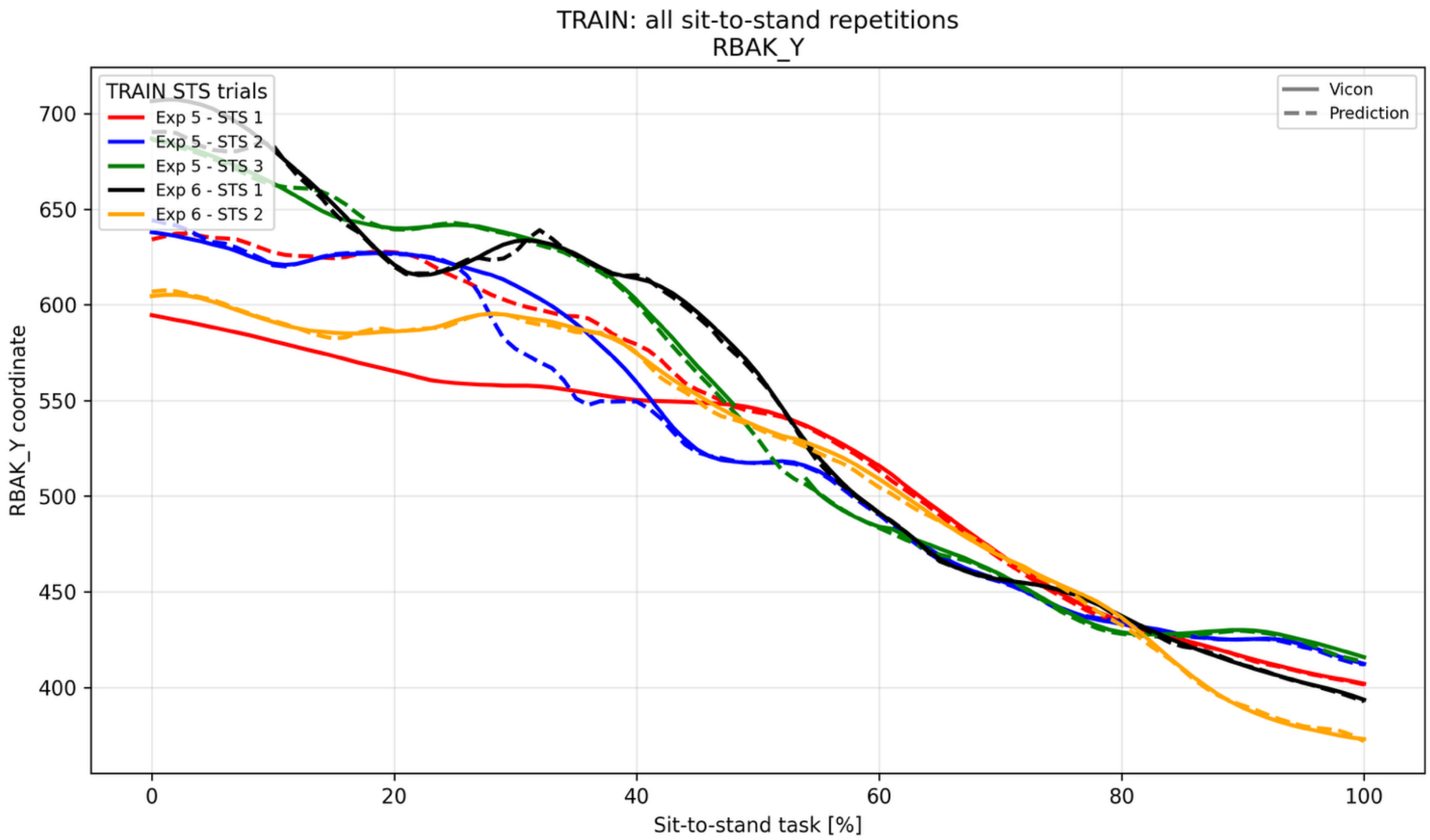

## Supplementary Figure 1c

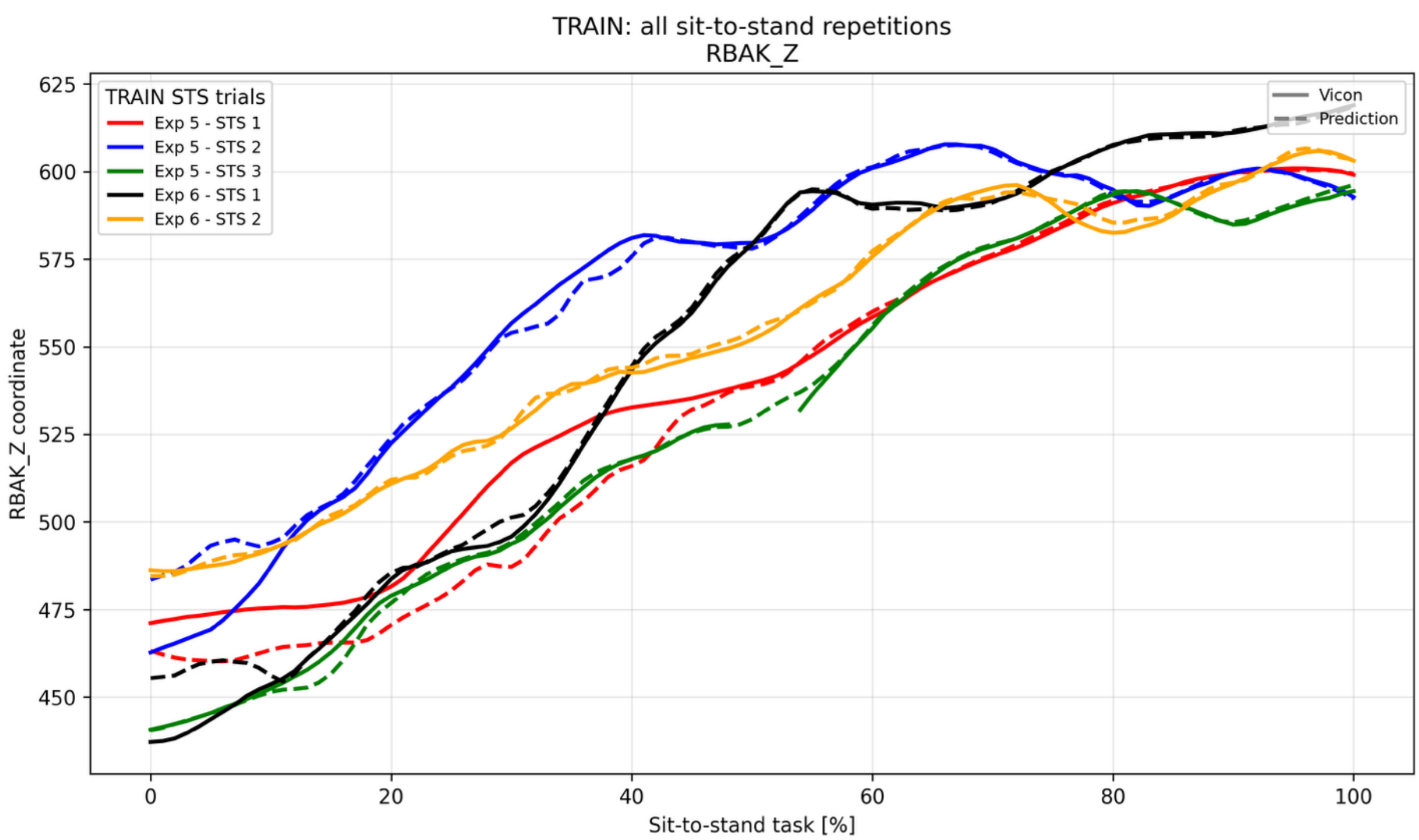

**Supplementary Fig. 2 | Training knee descriptors in the typically developing case.**

a–c, Sagittal-, frontal-, and transverse-plane projected thigh-shank descriptors in the aggregate training panels labelled left knee. Four repetitions are shown, with solid reference and dashed predicted curves. These aggregate panels are not presented as a same-limb comparison with the right-sided test panels in main Fig. 4. n = 4 repetitions from one participant.

## Supplementary Figure 2a

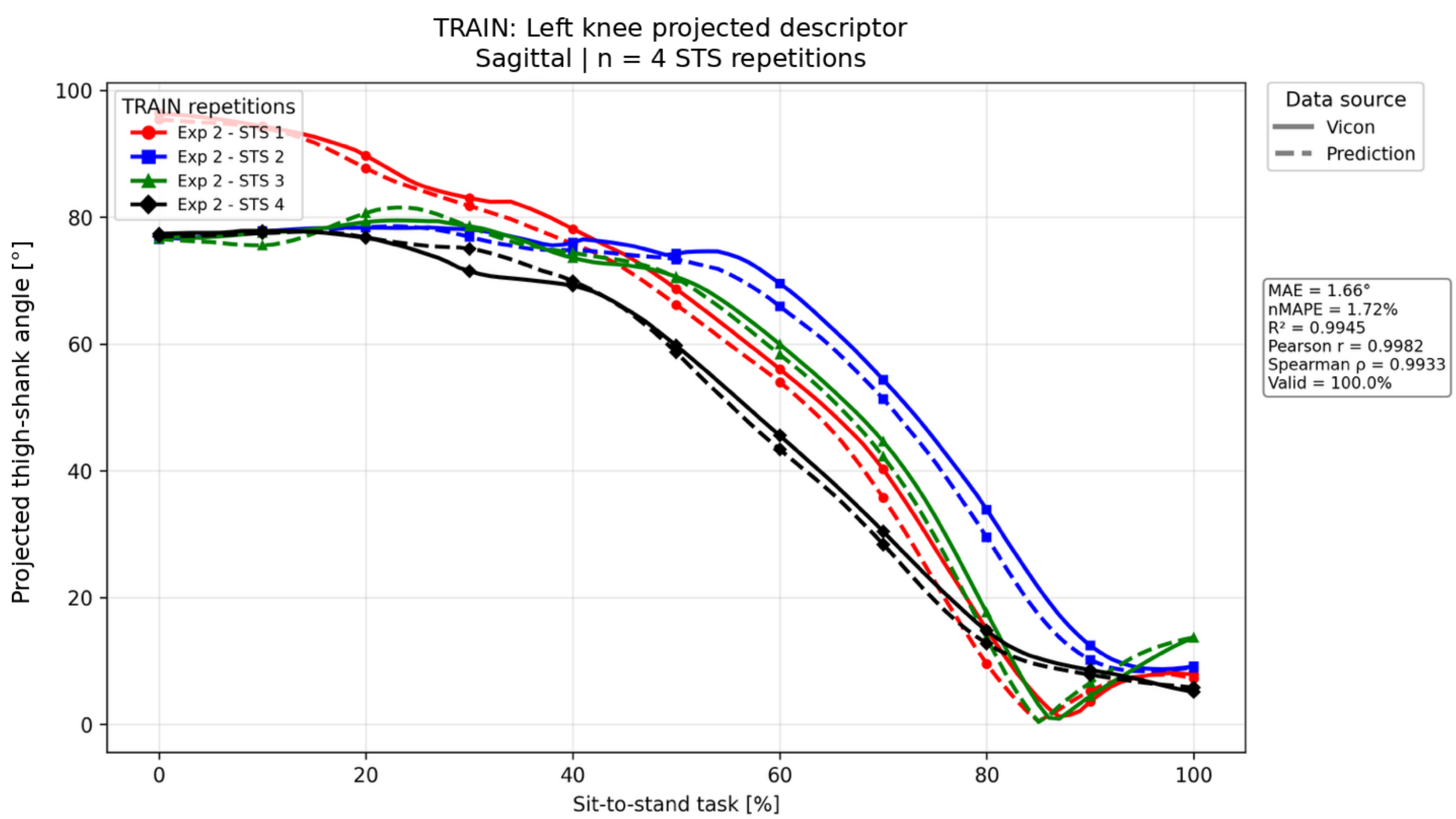

## Supplementary Figure 2b

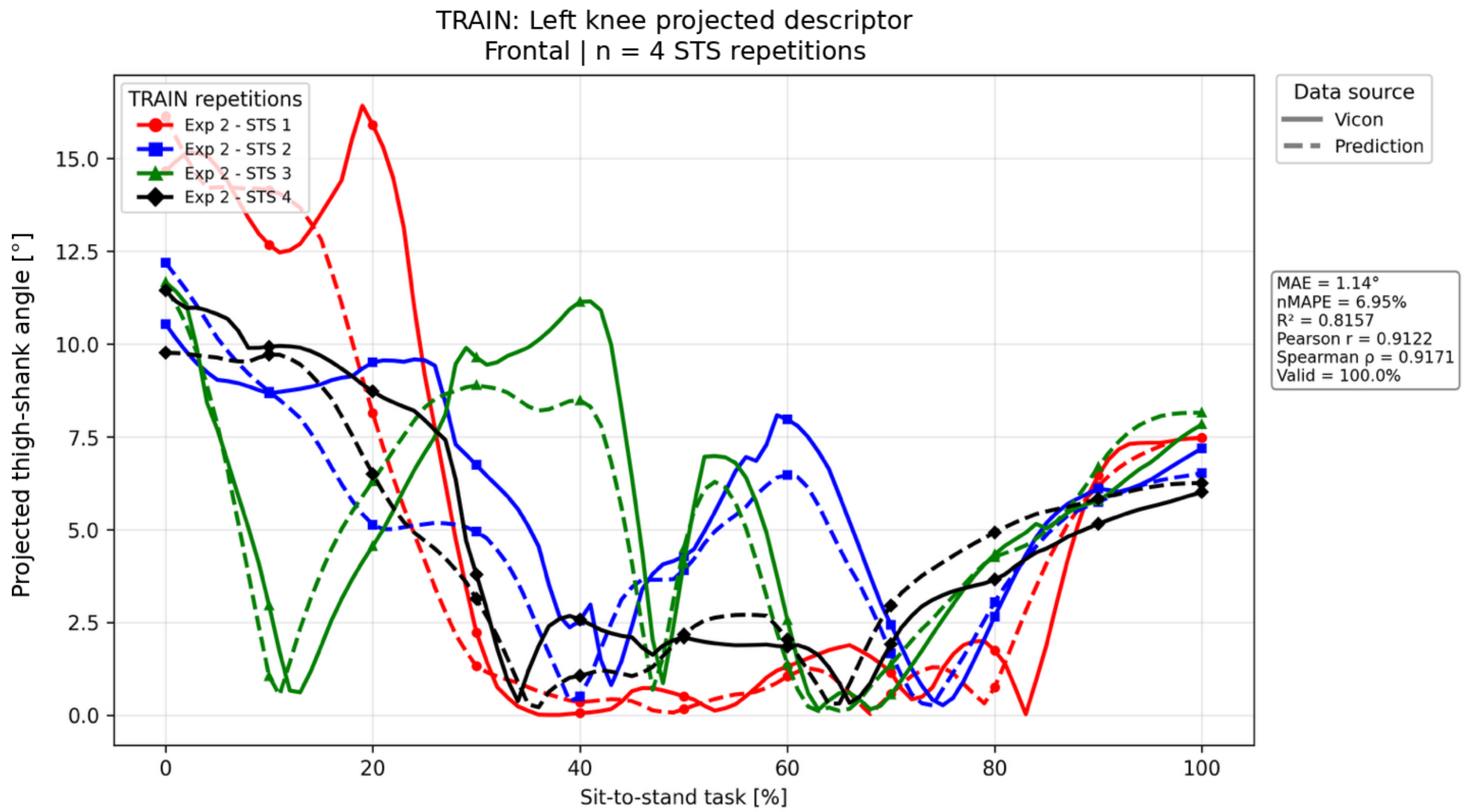

## Supplementary Figure 2c

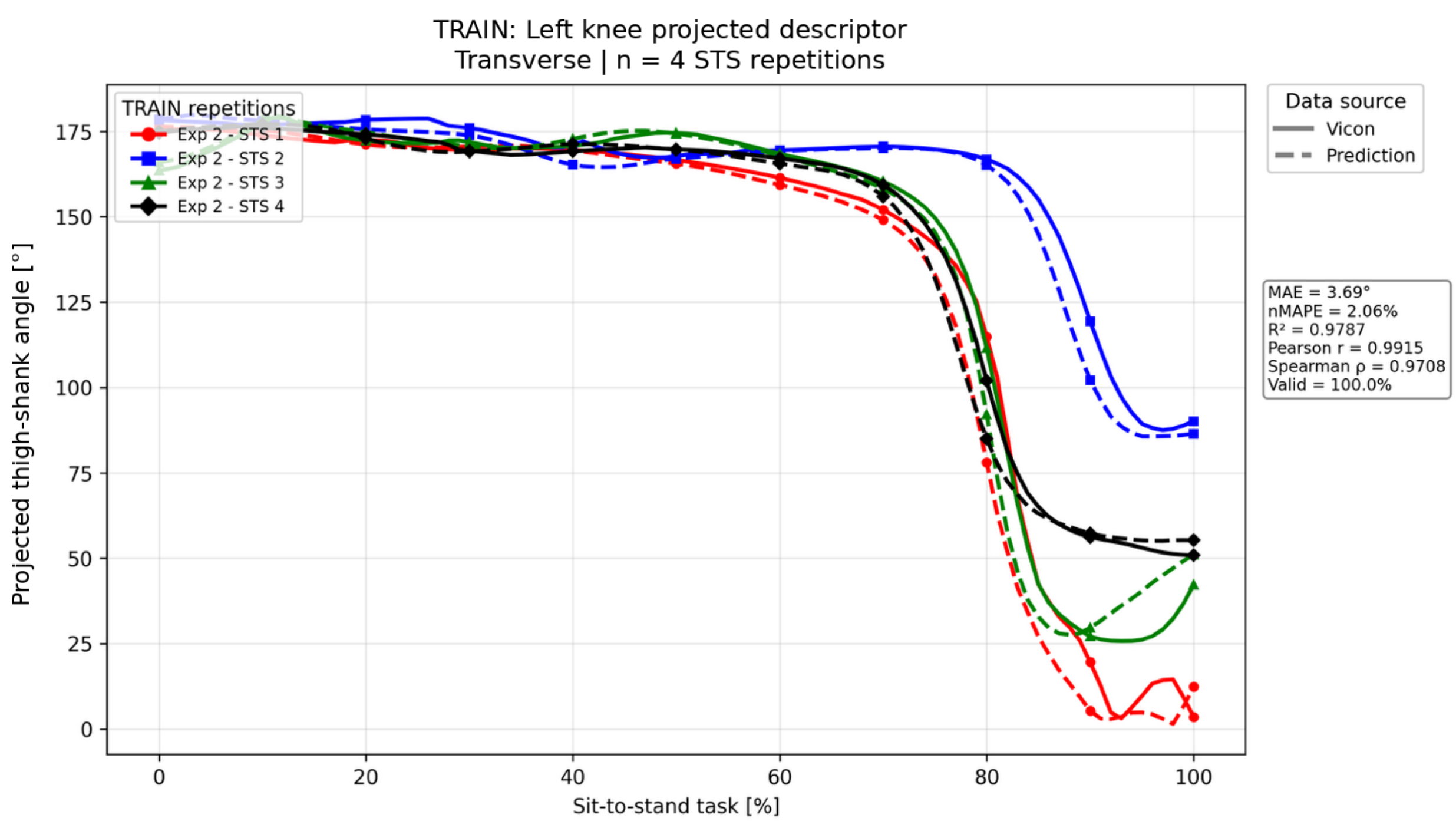

**Supplementary Fig. 3 | Training pelvis-relative trunk descriptors in the typically developing case.**

a–c, Sagittal, frontal, and transverse trunk descriptors for four training repetitions. Solid curves indicate the marker-based reference and dashed curves the predictions. These outcomes are distinct from the back-marker position and the knee descriptors. n = 4 repetitions from one participant.

## Supplementary Figure 3a

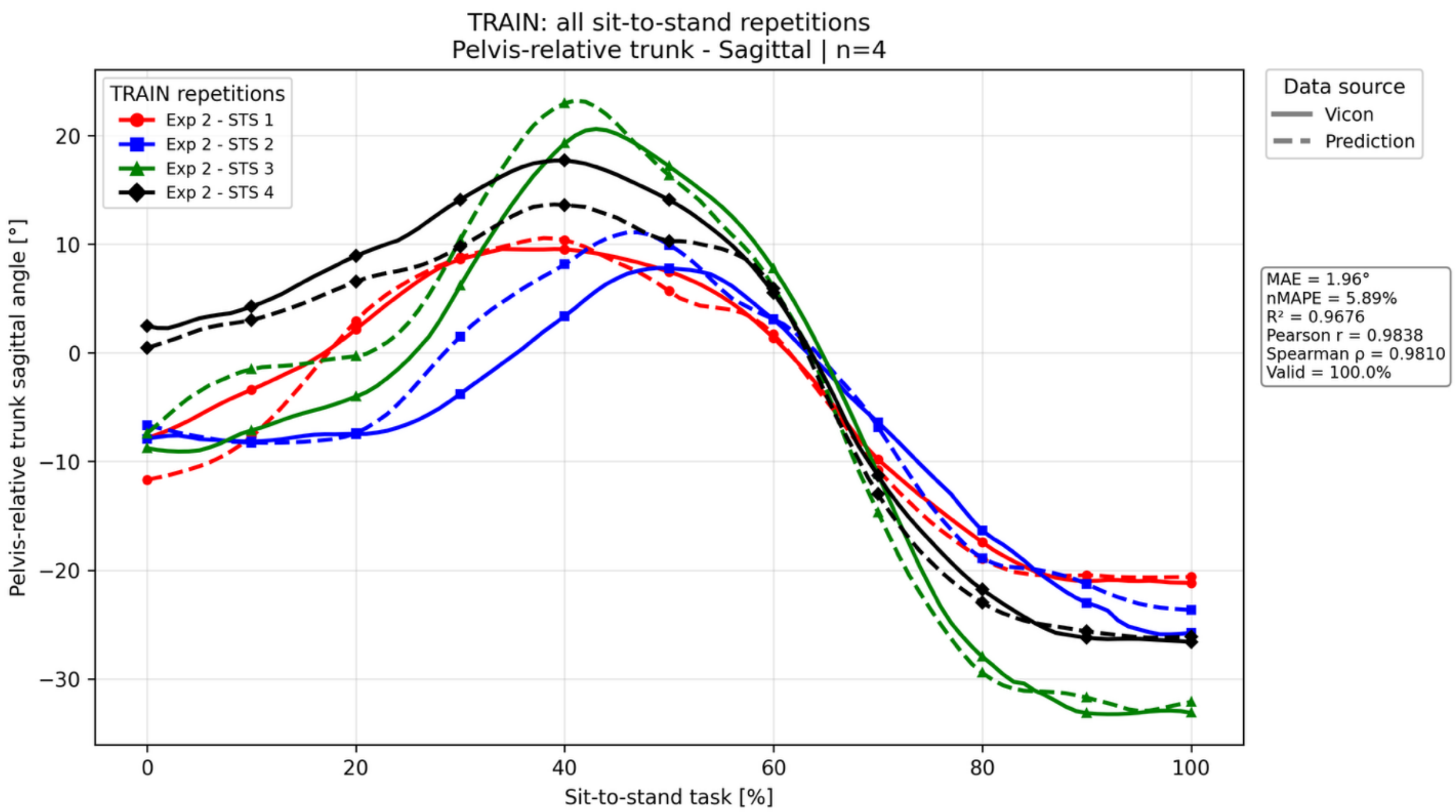

## Supplementary Figure 3b

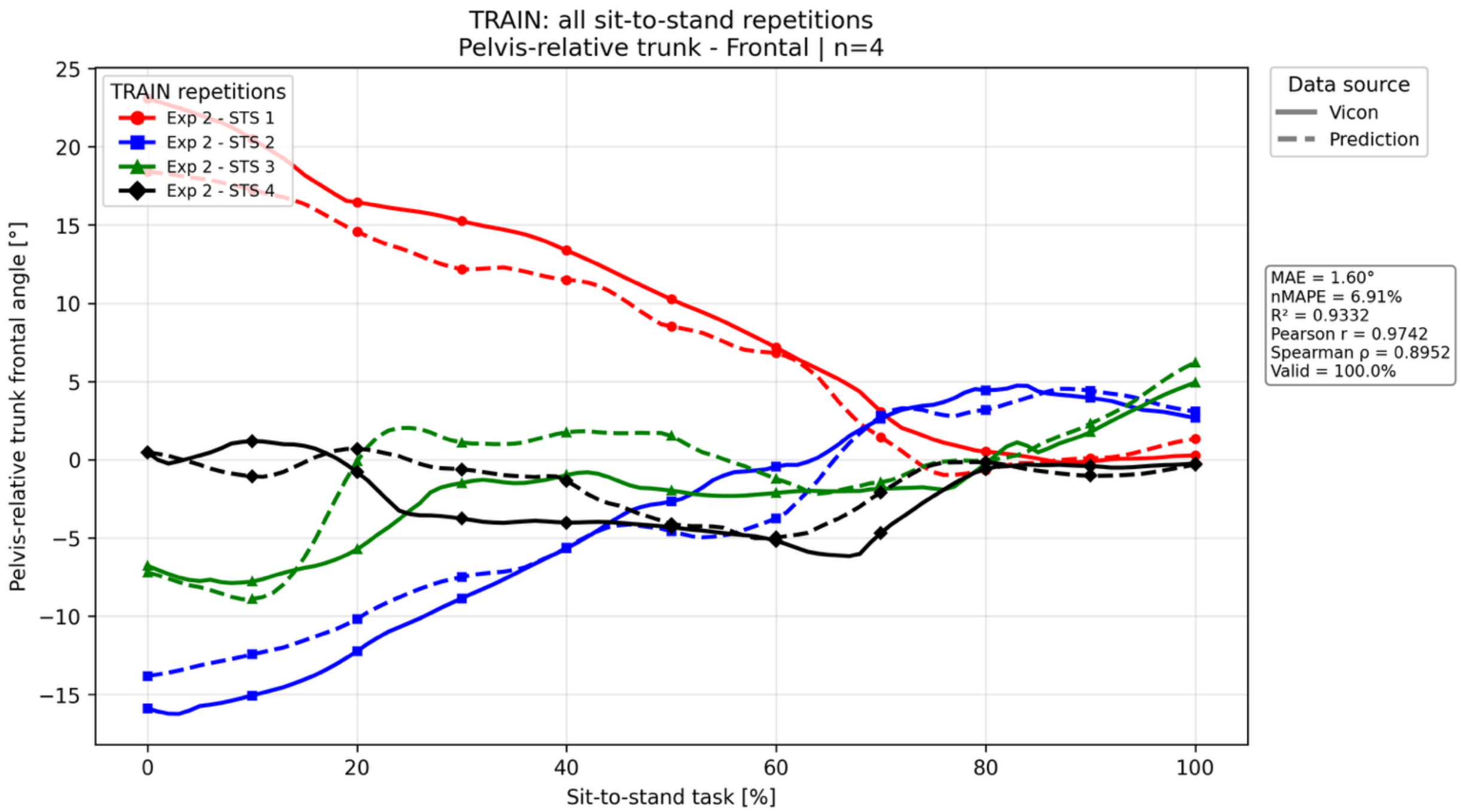

## Supplementary Figure 3c

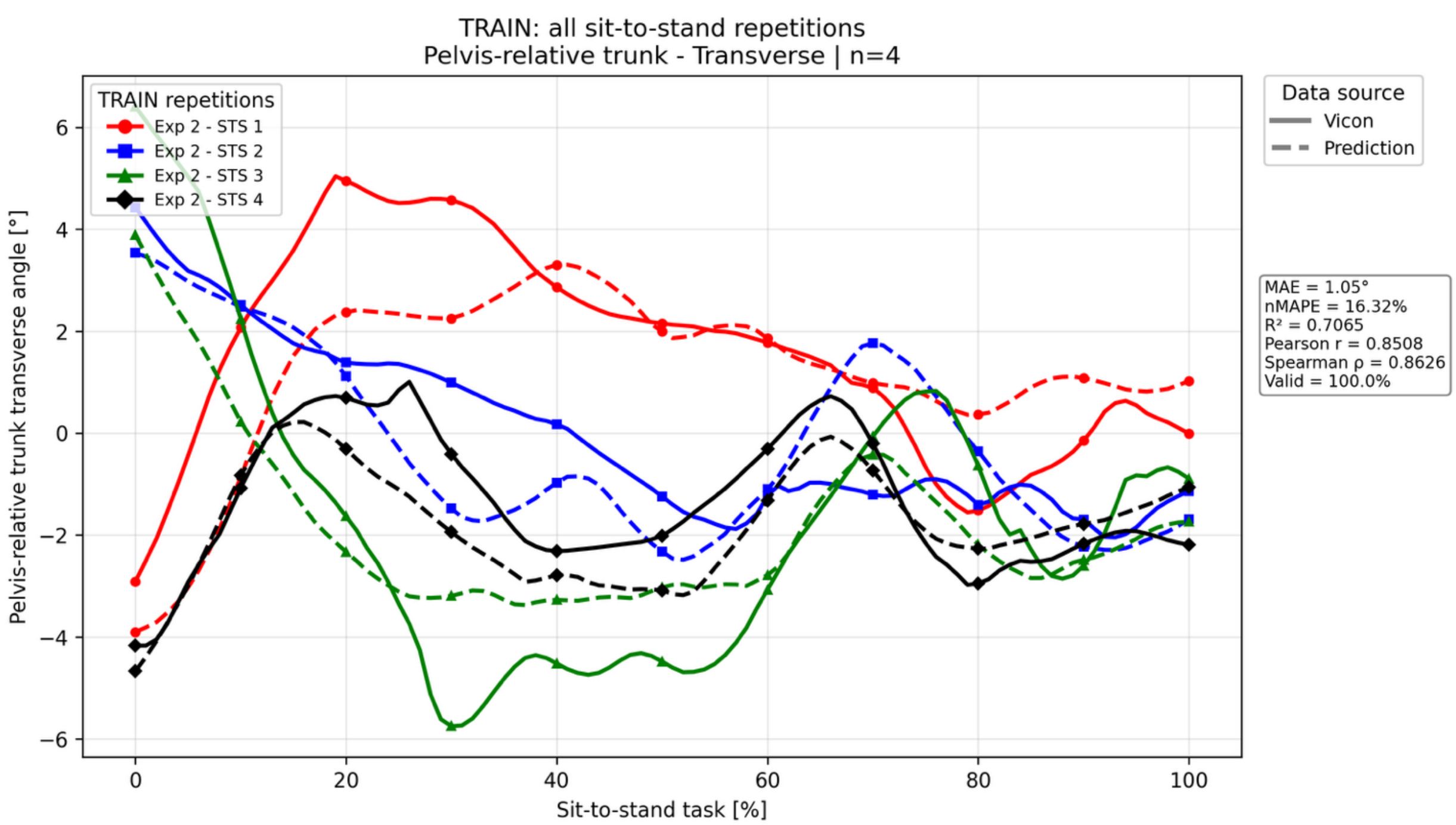

**Supplementary Fig. 4 | Training hip-knee cyclograms in the cerebral palsy case.**
Right-sided hip and knee flexion-proxy cyclograms for five training repetitions. Solid and dashed curves show reference and prediction. The summary reports hip MAE 0.53°, knee MAE 0.73°, paired-angle discrepancy 0.86°, and paired-data availability of 62.4%. Interpretation is limited to the samples entering the analysis, not all frames or all occlusion intervals. n = 5 repetitions from one participant.

**Supplementary Figure 4**

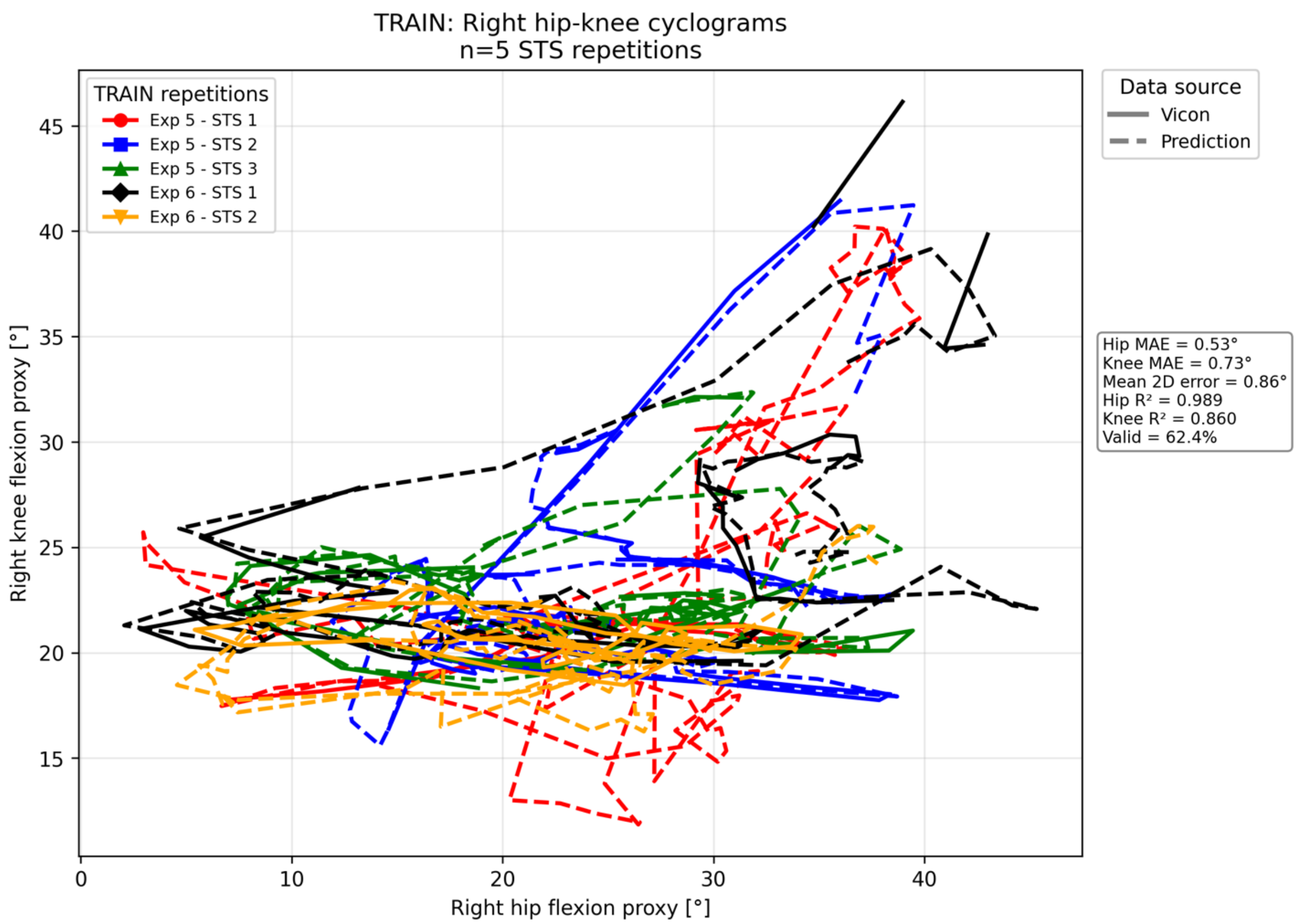

**Supplementary Fig. 5 | Training hip-knee cyclograms in the typically developing case.** Left-sided hip and knee flexion-proxy cyclograms for four training repetitions, with solid reference and dashed predicted curves. The summary reports hip MAE 1.96°, knee MAE 1.59°, paired-angle discrepancy 2.67°, and 100% paired-data availability. n = 4 repetitions from one participant.

**Supplementary Figure 5**

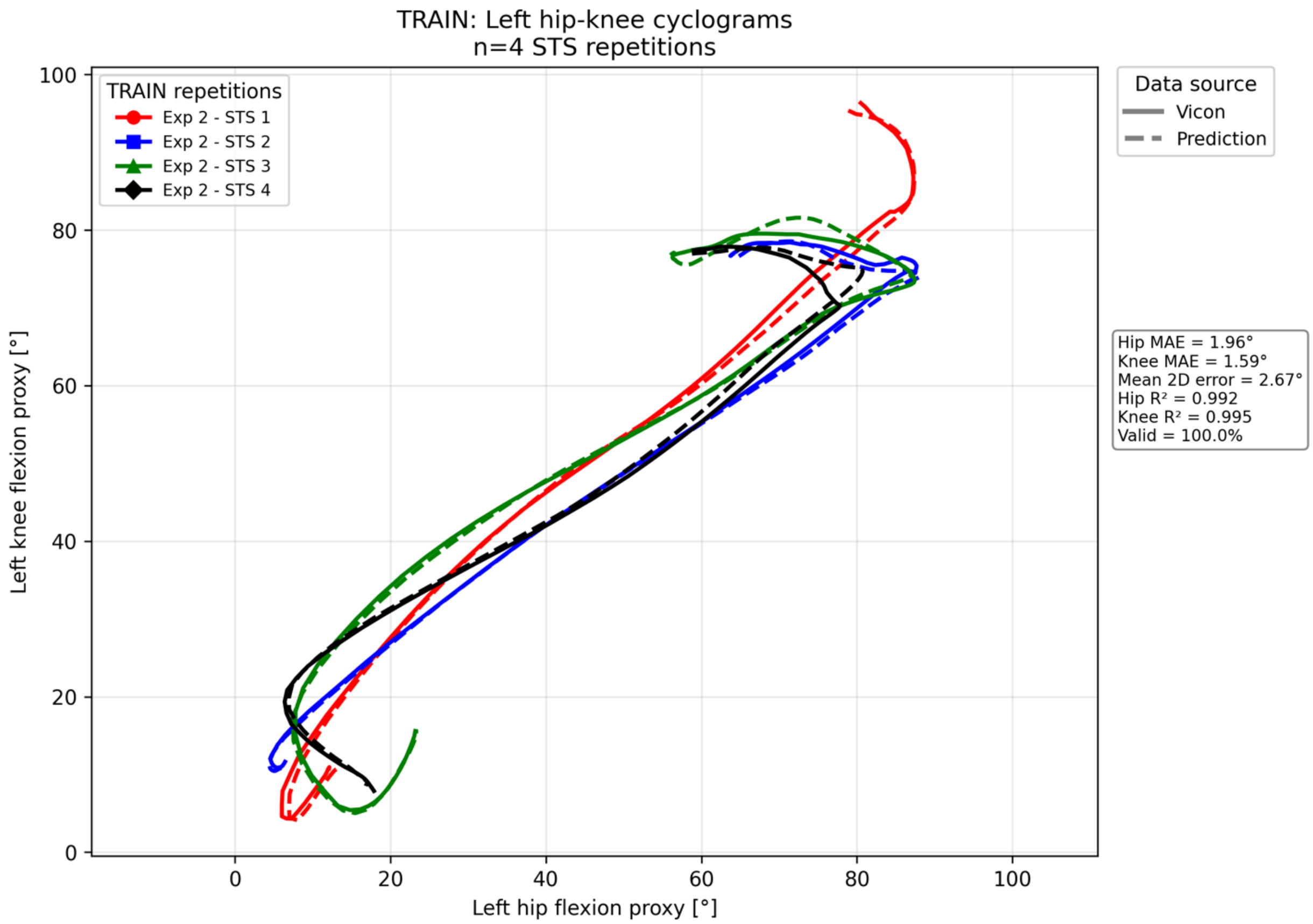

**Supplementary Fig. 6 | Training knee-marker reconstruction during multidirectional hopping.**

LKNE Z-coordinate trajectories for 16 individual training hops, with solid reference and dashed predicted curves. Individual hops are components of the recorded sequence rather than independent participants. The summary reports MAE 5.984 mm and $R^2$ 0.9823. The horizontal axis denotes normalized hop progression. n = 16 hops from one participant.

**Supplementary Figure 6**

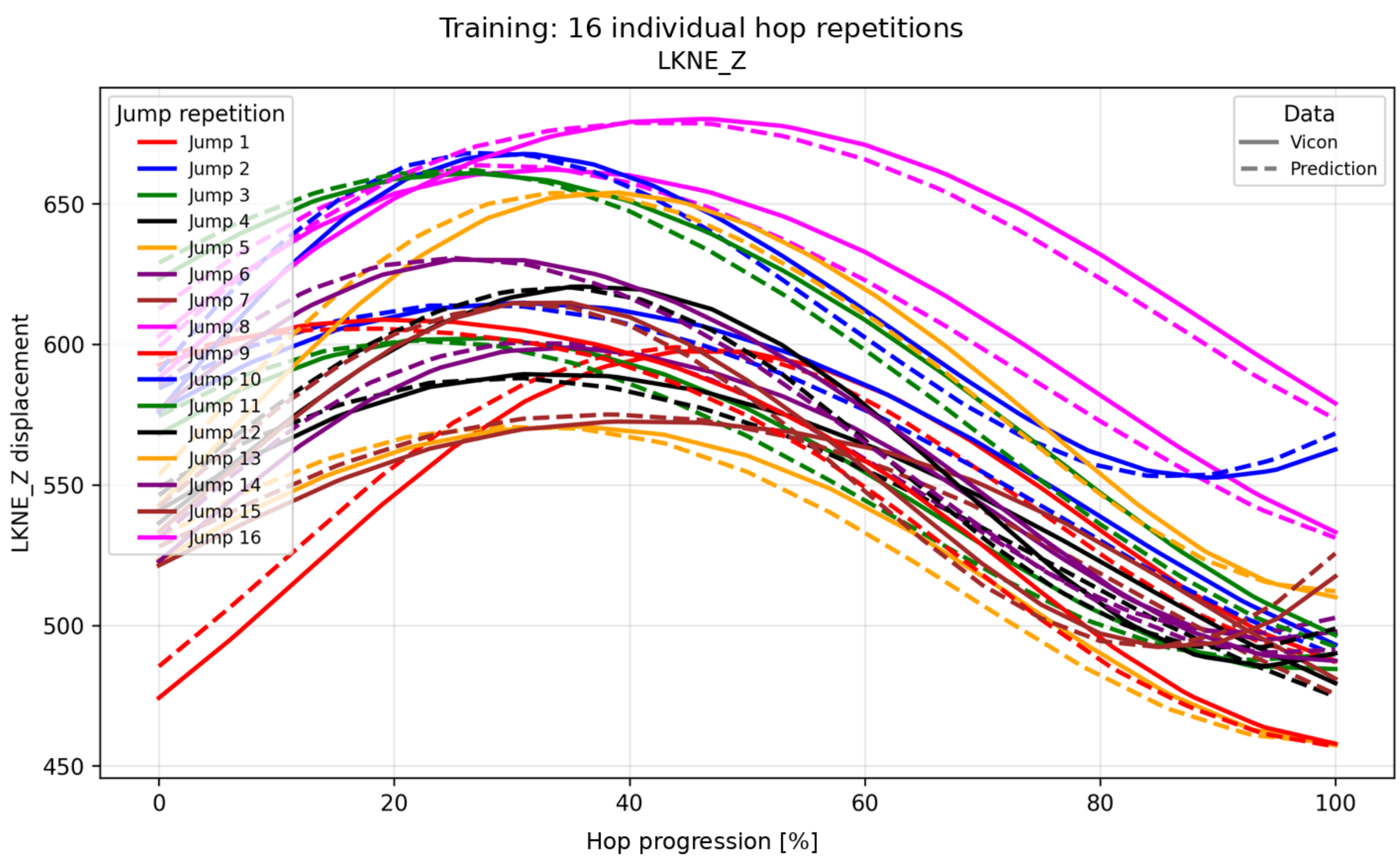

**Supplementary Fig. 7 | Sample Flight time identification in the CUBE use case.**

a, Sample train flight status 1. b, Sample train flight status 2. c, Sample test flight status 1. d, Sample test flight status 2. Flight status was post processed from the gold standard data and prediction data using the same algorithm. The ground level was identified from the minimum Z coordinate of the toe marker along the selected jump sequence. The "in-flight" status is deemed to be a Z coordinate of the toe marker with a value of 1 mm above the identified ground level.

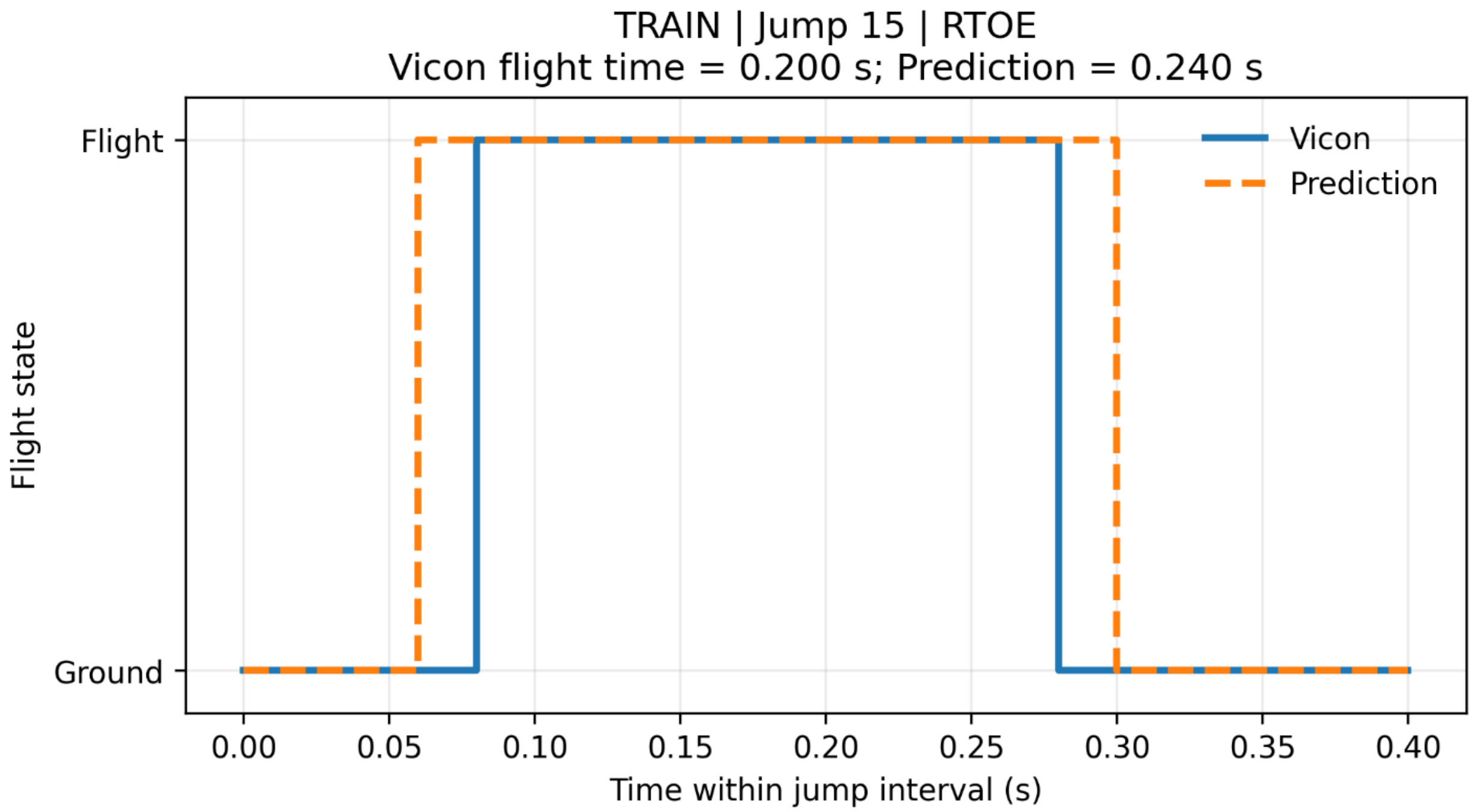

**Supplementary Figure 7(a)** Sample train flight status 1.

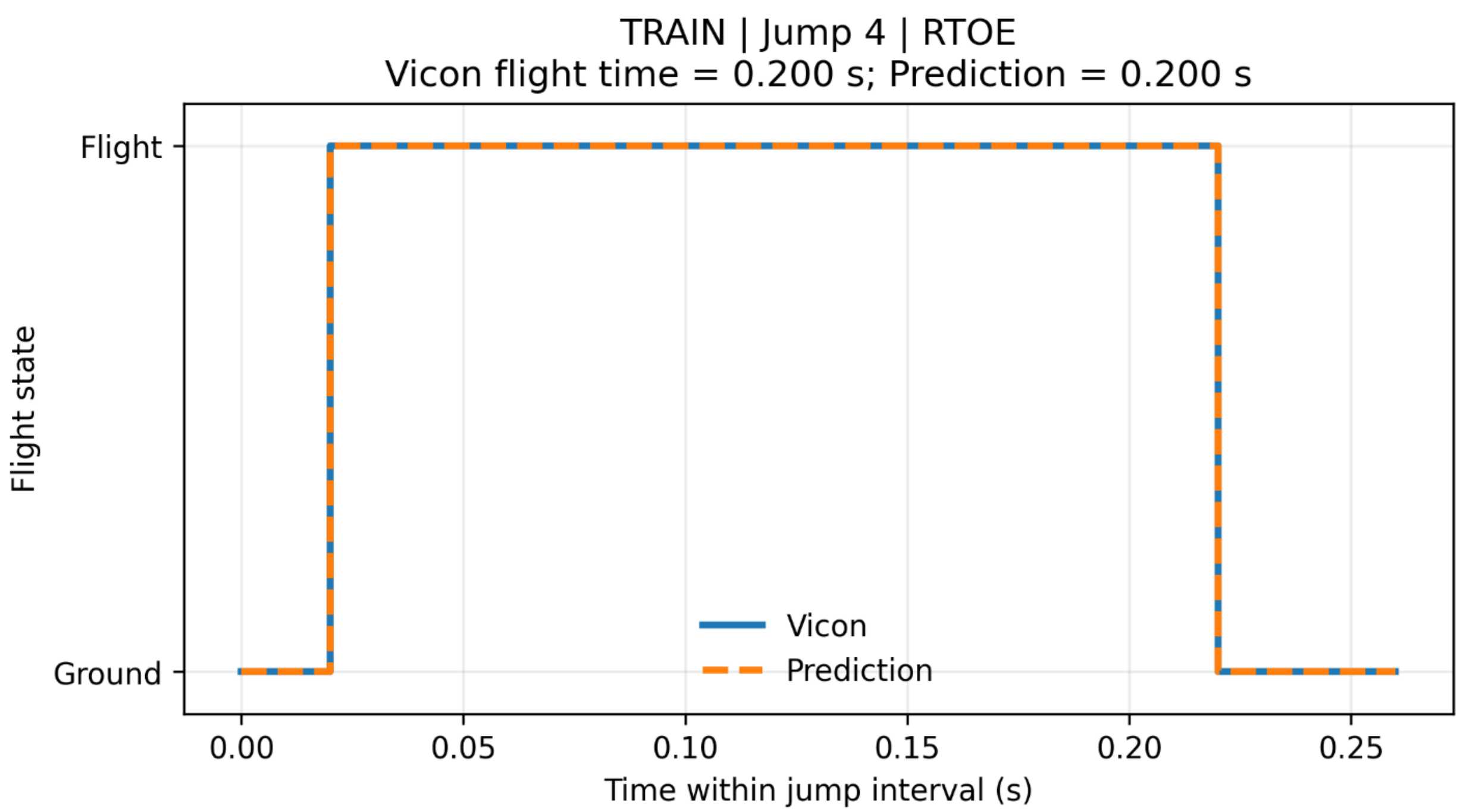

**Supplementary Figure 7(b)** Sample train flight status 2.

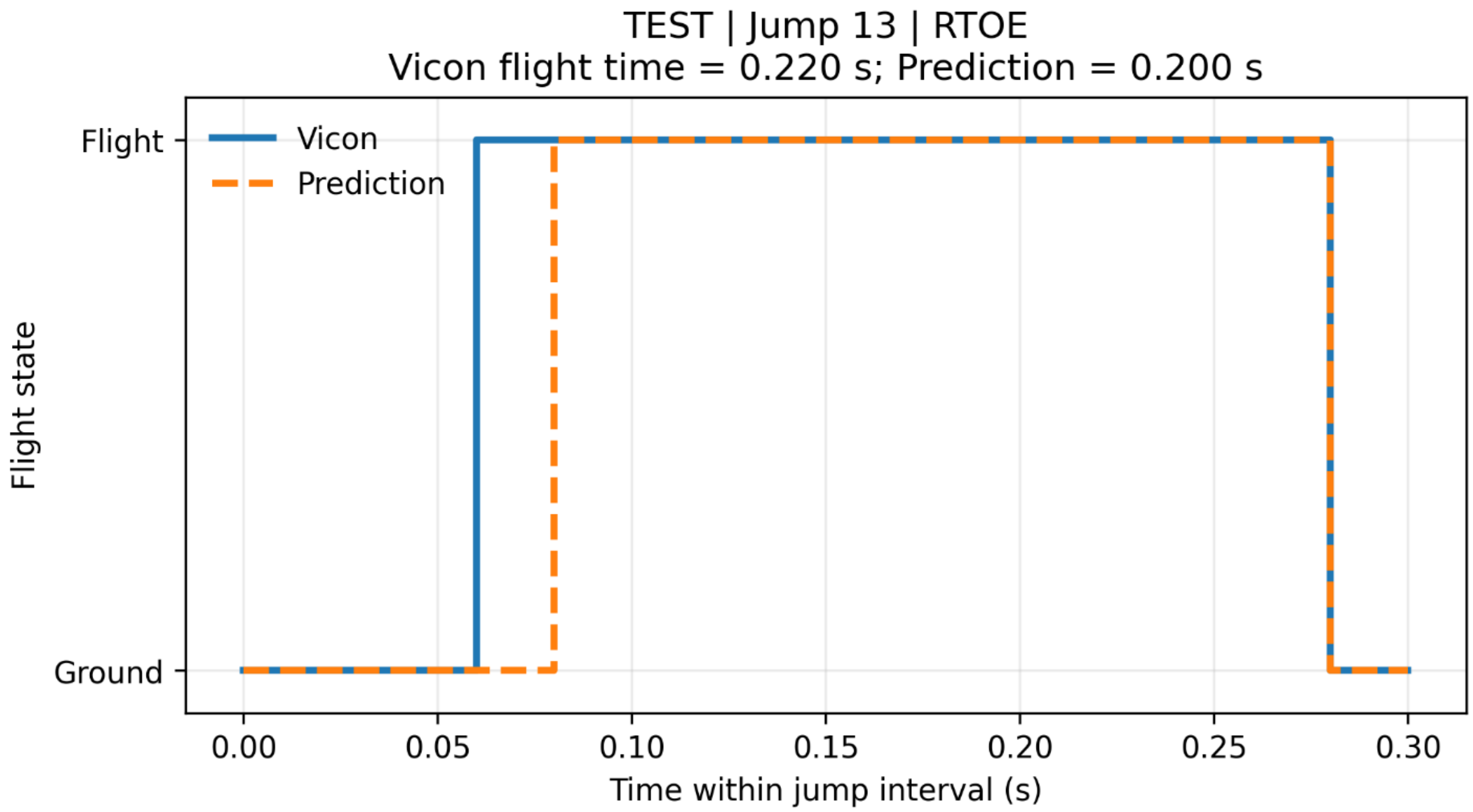

**Supplementary Figure 7(c)** Sample test flight status 1.

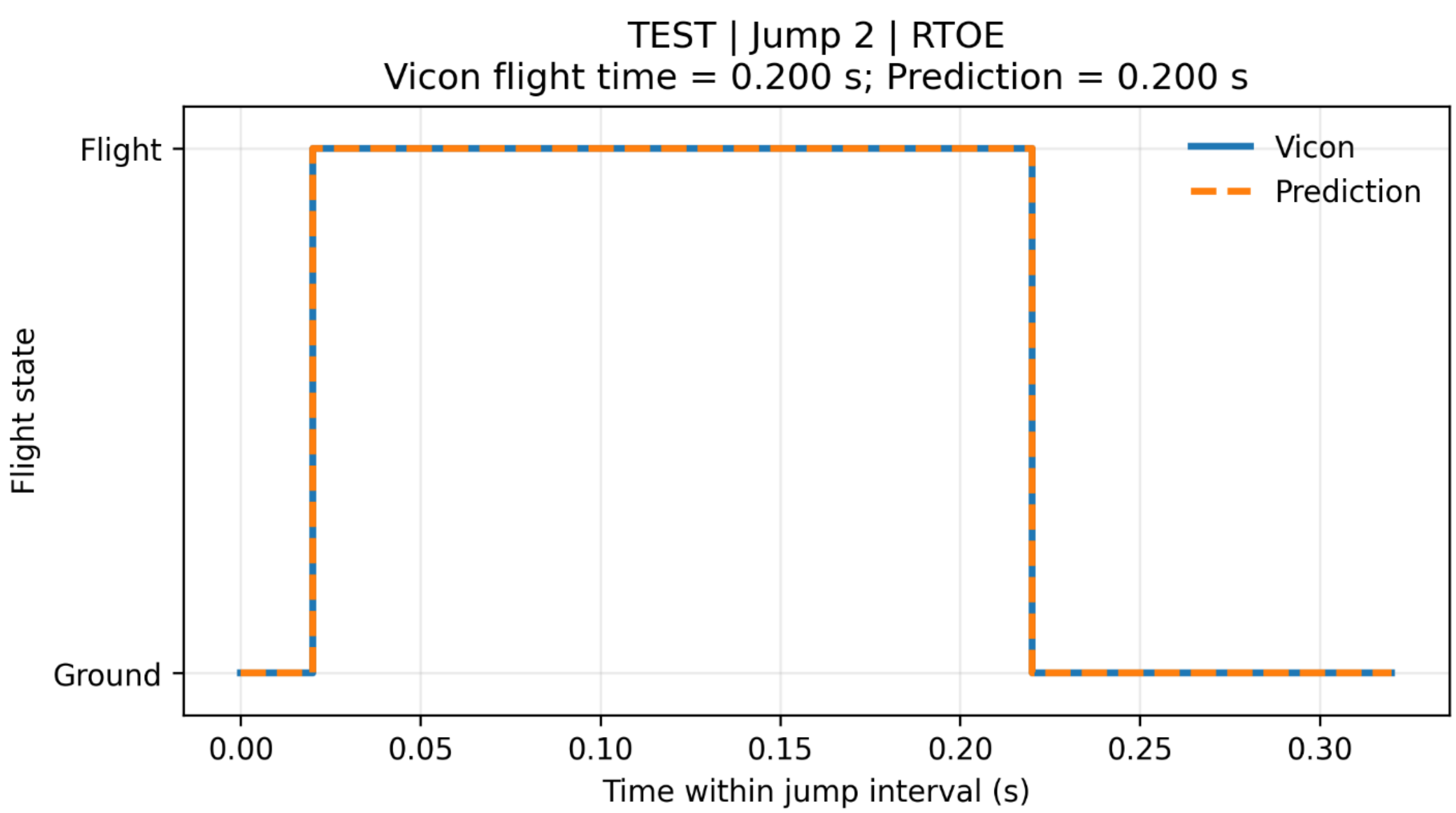

**Supplementary Figure 7(d)** Sample train flight status 2.

**Supplementary Fig. 8 | Convolutional encoder, image decoder, and coordinate regressor.**

a, View-specific convolutional encoder in the supplied cerebral palsy configuration. b, Decoder reconstructing the six-channel image representation. c, Feedforward multilayer perceptron mapping the concatenated representation to coordinate outputs. Conv2D denotes two-dimensional convolution; Conv2DTranspose denotes transposed two-dimensional convolution; ReLU denotes rectified linear activation. Tensor labels are reproduced from the supplied model exports, and None denotes the variable batch dimension. Retained SVD rank is distinct from the dimensionality of the encoder output.

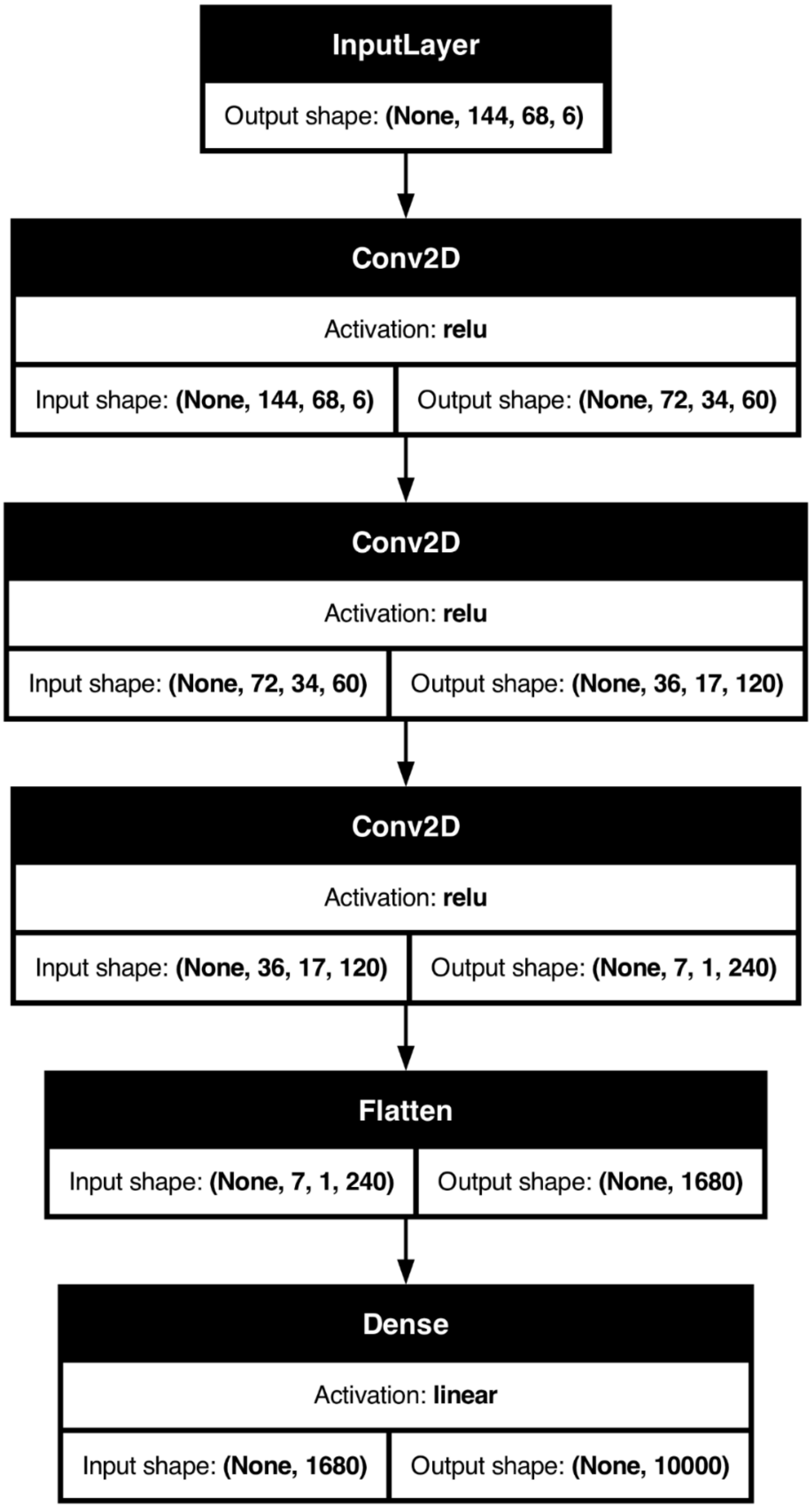

**Supplementary Figure 8(a)**

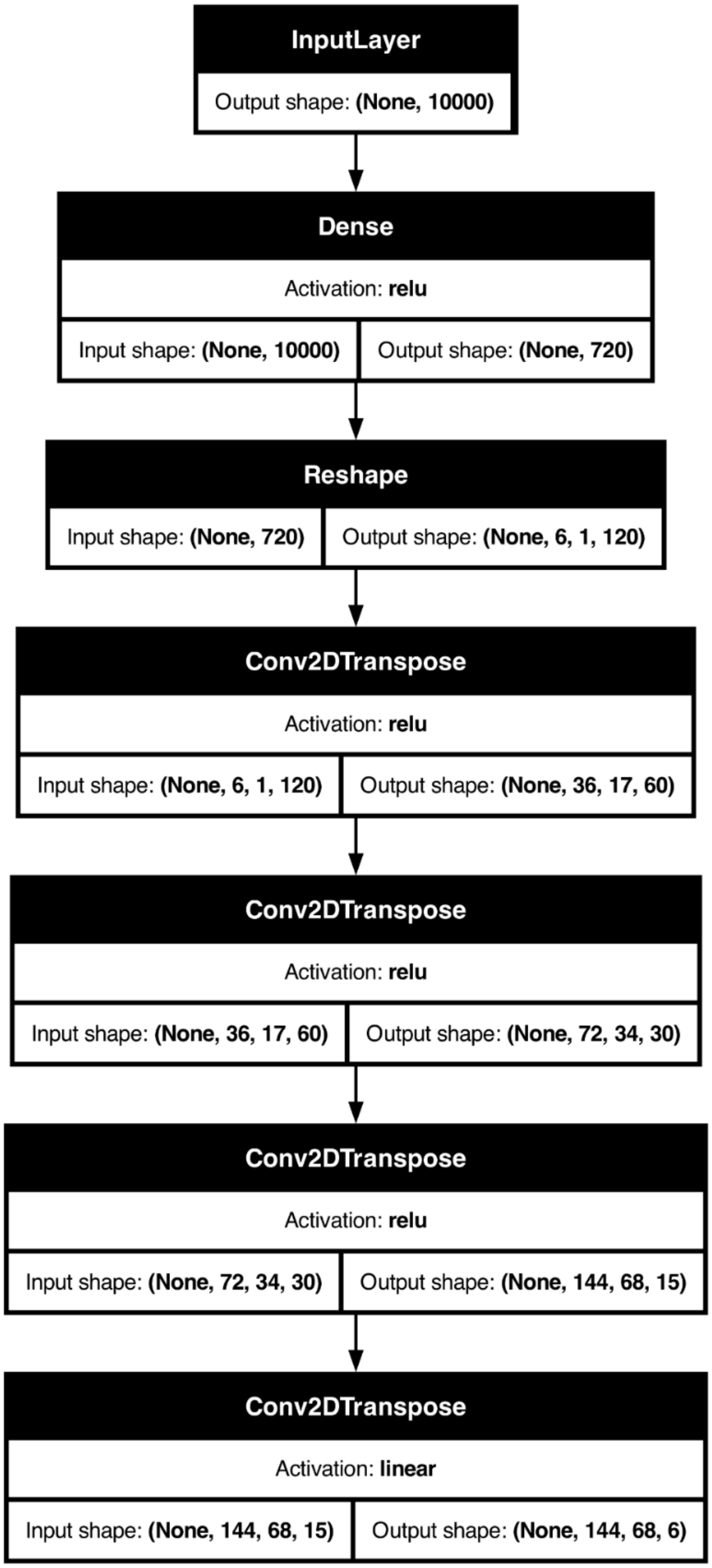

**Supplementary Figure 8(b)**

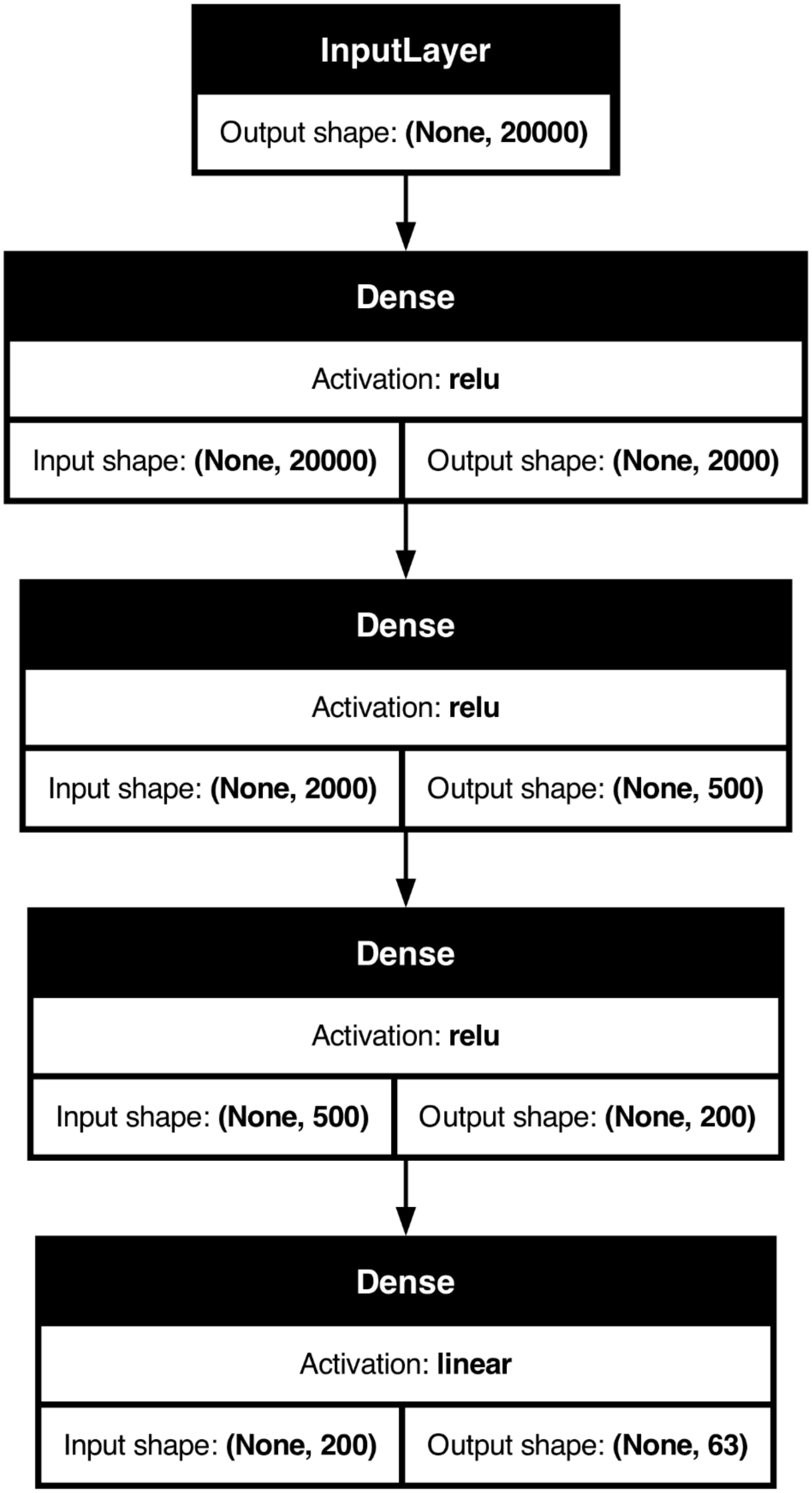

**Supplementary Figure 8(c)**